\documentclass[12pt]{article}

\usepackage{newtxtext,newtxmath}

\usepackage{graphicx}

\usepackage[letterpaper,margin=1in]{geometry}

\renewenvironment{abstract}
	{\quotation}
	{\endquotation}

\date{}

\makeatletter
\renewcommand{\fnum@figure}{\textbf{Figure \thefigure}}
\renewcommand{\fnum@table}{\textbf{Table \thetable}}
\makeatother

\usepackage{scicite}

\usepackage{url}

\newcommand{\jgr}{Journal of Geophysical Research}
\newcommand{\icarus}{Icarus}
\newcommand{\psj}{The Planetary Science Journal}
\newcommand{\aj}{The Astronomical Journal}
\newcommand{\apj}{The Astrophysical Journal}
\newcommand{\apjl}{The Astrophysical Journal Letters}
\newcommand{\aap}{Astronomy \& Astrophysics}
\newcommand{\nat}{Nature}
\newcommand{\mnras}{Monthly Notices of the Royal Astronomical Society}
\newcommand{\grl}{Geophysical Research Letters}
\newcommand{\jqsrt}{Journal of Quantitative Spectroscopy and Radiative Transfer}

\def\scititle{
	Tracing the Inner Edge of the Habitable Zone with Sulfur Chemistry
}
\title{\bfseries \boldmath \scititle}

\author{
	Sean~Jordan$^{1,2\ast}$,
	Oliver~Shorttle$^{2,3}$,
	Paul~B.~Rimmer$^{4}$\and
	\small$^{1}$ETH Zurich, Institute for Particle and Astrophysics, Wolfgang-Pauli-Strasse 27, CH-8093 Zurich, Switzerland.\and
	\small$^{2}$Institute of Astronomy, University of Cambridge, Cambridge \& CB3 0HA, United Kingdom.\and
	\small$^{3}$Department of Earth Sciences, University of Cambridge, Cambridge \& CB2 3EQ, United Kingdom.\and
	\small$^{4}$Cavendish Laboratory, University of Cambridge, Cambridge \& CB3 0HE, United Kingdom.\and
	\small$^\ast$Corresponding author. Email: jordans@ethz.ch
}

\begin{document} 

\maketitle

\begin{abstract} \bfseries \boldmath
The circumstellar liquid-water habitable zone guides our search for potentially inhabited exoplanets, but remains observationally untested.  We show that the inner edge of the habitable zone can now be mapped among exoplanets using their \textit{lack} of surface water, which, unlike the \textit{presence} of water, can be unambiguously revealed by atmospheric sulfur species. Using coupled climate-chemistry modelling we find that the observability of sulfur-gases on exoplanets depends critically on the ultraviolet (UV) flux of their host star, a property with wide variation: most M-dwarfs have a low UV flux and thereby allow the detection of sulfur-gases as a tracer of dry planetary surfaces; however, the UV flux of Trappist-1 may be too high for sulfur to disambiguate uninhabitable from habitable surfaces on any of its planets.  We generalise this result to show how a population-level search for sulfur-chemistry on M-dwarf planets can be used to empirically define the Habitable Zone in the near-future.
\end{abstract}


\section*{Introduction}
\label{sec:into}

\noindent Remote observation of exoplanet atmospheres offers the opportunity to test the origin and prevalence of life in the universe. In order to perform this test, we must be able to distinguish between three possible cases: uninhabitable planets, habitable but lifeless planets, and habitable planets that are inhabited by life. Previous research has largely focused on our ability to detect the latter of these three cases, planets inhabited by life, via the detection of gases associated with biological activity. This implicitly takes for granted the idea that we can distinguish between the former two cases: uninhabitable and habitable planets. This, however, remains an unsolved problem.

Venus and the Earth provide us with local paradigms of uninhabitable versus habitable planets within the Solar System. Venus’s surface is heated to 735\,K due to the greenhouse effect of its CO$_2$-dominated atmosphere and global cloud cover, while the Earth has maintained surface temperatures suitable for liquid water over geological timescales. With respect to exoplanets, the Earth represents the paradigm of surface habitability and Venus represents the paradigm of surface hostility and the death of habitability. The distinct climatic fates of these solar system planets have inspired the concept of a Habitable Zone (HZ) around stars more generally: bounded at its inner edge by the Runaway Greenhouse limit where an Earth-like planet would lose its liquid water oceans and evolve into a Venus-like state.

Estimates of the inner edge of the Habitable Zone depend heavily on modelling assumptions and whether or not a planet has evolved from a `hot start’ or a `cold start’. Models of slowly rotating planets initialised with surface water oceans (a `cold start') can maintain habitable conditions well inside the traditional `Venus Zone' \cite{Kane2014} of a star via a cloud-climate feedback \cite{Yang2013,WayDelGenio2020}. Alternatively, models of planets initialised with a `hot start', due to the accretional energy that must be released during the planet's formation, require lower instellation fluxes than the traditional Runaway Greenhouse limit in order to initially condense surface water oceans, a limit known as the `Water Condensation Zone' (WCZ) \cite{Turbet2023}. The physical implication of these contrasting results is that climate hysteresis pushes the HZ inner edge further out if planets have formed with a hot start \textit{in situ} \cite{Turbet2023}.

The uncertainties on the HZ inner edge are greatly magnified for planets orbiting M-dwarfs, which are the systems with rocky planets that current observational facilities are best able to prospect for life. Tidal-locking of planets throughout the Venus Zones of M-dwarfs ensures that such planets are slowly rotating and therefore will achieve long-term habitability if they attained a `cold start' from mechanisms such as planetary migration, however this climate stability would not be achieved for planets with even moderately faster rotational periods, such as those in a spin-orbit resonance \cite{Yang2013}. The water-condensation limit for the `hot start' scenario on the other hand is exacerbated by the extended super-luminous pre-Main Sequence phase of an M-dwarf’s stellar evolution \cite{Baraffe2002, Ramirez2014}: a planet that formed in the present day Habitable Zone of an M-dwarf will have been exposed to orders of magnitude greater instellation fluxes for up to several 100s of Myrs in its early evolution, capable of desiccating the planet and leaving behind a reservoir of O$_2$ from water photolysis and hydrogen escape. This has been hypothesised to result in a potential false-positive detection of planets with Earth-like atmospheres in the HZ of M-dwarfs \cite{LugerBarnes2015}, that could instead be hot and desiccated, with O$_2$-rich relic atmospheres.

Now that we have entered the era of atmospheric characterisation of rocky planet atmospheres with the James Webb Space Telescope, the critical next step in testing the origin and prevalence of life on nearby exoplanets lies in mapping the location of the HZ inner edge around M-dwarf host stars. The primary challenge to this task is in breaking the degeneracy between observations of Earth-like and Venus-like planets at relatively low signal-to-noise, which occurs from their high mean molecular weight atmospheres and small planet-star size ratios compared to gas giants or sub-Neptunes. Observational modelling of planets with Venus-like atmospheres has revealed that planets with wide cloud coverage are difficult to distinguish from planets with thin atmospheres: without the \textit{a priori} knowledge that an atmosphere is cloudy and Venus-like, the statistical retrieval of atmospheric properties from observations can favour a clear-sky atmosphere with a retrieved surface pressure equal to the true cloud-top pressure \cite{Konrad2023}.

Diagnostic observational indicators of Venus-like versus Earth-like paradigms therefore need to exist in the upper layers of an exoplanet’s atmosphere to be observationally accessible. Observational identification of an Earth-like atmosphere relies primarily on the strong ozone feature resulting from Earth-like levels of oxygen \cite{Barstow2016}. This feature, however, cannot sufficiently diagnose either habitable or inhabited planets for three reasons: (a) the link between ozone abundance and oxygen abundance throughout Earth history is itself ambiguous [e.g., \cite{Cooke2022}]; (b) ozone is observed in greater abundance on Venus than the upper limits on the oxygen abundance would imply, suggesting that there are unknown mechanisms for enriching the ratio of O$_3$:O$_2$ in planetary atmospheres \cite{Montmessin2011, Marcq2019}; (c) oxygenated atmospheres may be produced on planets as a result of passing through the Runaway Greenhouse transition \cite{Wordsworth2014,Krissansen-Totton2022}.

In the absence of unique chemical identifiers of an Earth-like atmosphere, positively identifying a Venus-like atmosphere, with a surface too hot to support liquid water, is therefore imperative for constraining the bounds of habitability amongst the rocky exoplanet population. Venus’s atmosphere is dominated by CO$_2$, whereas in the Earth’s atmosphere CO$_2$ exists in only trace abundance. In transmission spectroscopy, a technique routinely applied by JWST, increasing the CO$_2$ mass fraction in an atmosphere decreases the scale height and compensates for the stronger CO$_2$ features in the spectrum, thus Earth-like and Venus-like atmospheres remain indistinguishable via this observation method unless the Rayleigh scattering slope can be well resolved \cite{BennekeSeager2012, Triaud2023}. In emission spectroscopy, the high pressure CO$_2$-dominated atmosphere of a Venus-twin would be more easily diagnosed on a short-period exoplanet compared to in transmission spectroscopy \cite{zieba2023}, however the detection of CO$_2$ on planets with lower pressure atmospheres and/or obscuring clouds and hazes remains ambiguously linked to the planet’s true climate and potential to support surface water \cite{LustigYaeger2019}. This is problematic at the low signal to noise JWST will attain in emission for the colder rocky planets we expect at the HZ inner edge \cite{LustigYaeger2019}. Such observations are also insensitive to the presence of other filler gases that may be invisible in emission spectroscopy, but alter the surface pressure, planetary climate and thereby habitability \cite{zieba2023}.

One chemical species that can potentially be used to unambiguously map between upper atmosphere observations and conditions at the surface of a rocky planet is sulfur dioxide (SO$_2$). Sulfur dioxide is abundant in Venus's atmosphere compared to the Earth's, and is expected to be scrubbed from habitable planet's atmospheres more generally via wet deposition \cite{Loftus2019} (i.e., dissolution into liquid water through rainout, sequestration in surface oceans, and precipitation into mineral phases). Since Venus lacks a hydrological cycle due to having lost its primordial water inventory, SO$_2$ can build up to significant abundances in its atmosphere. This is observed to be true in the lower atmosphere of Venus where SO$_2$ is the third most abundant atmospheric gas, however it is not true in the upper atmosphere due to the efficient photochemical depletion of SO$_2$ by Solar irradiation, resulting in H$_2$SO$_4$ cloud-formation. SO$_2$ would therefore not be observed on a Venus-twin around a G-dwarf host star \cite{Konrad2023}, despite Venus's lack of a hydrological cycle and surface water oceans.

\begin{figure}[ht!]
    \centering
    \includegraphics[width=\textwidth]{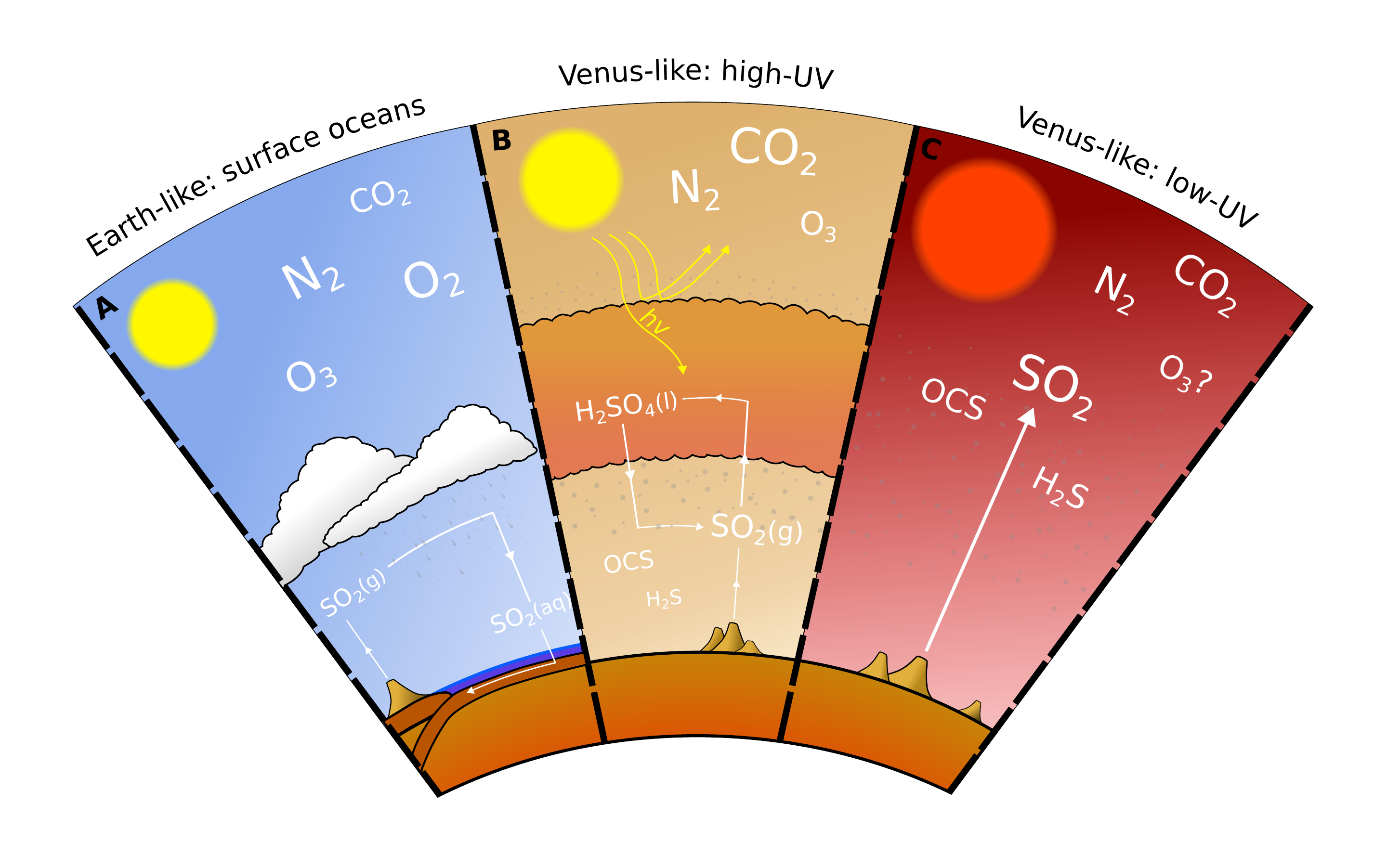}
    \caption{\textbf{Cartoon depiction of the sulfur-cycle in different planetary regimes.} In the Earth-like case (A), atmospheric SO$_2$ is scrubbed from the atmosphere by wet deposition \cite{Loftus2019}. In the Venus-like case with high-UV irradiation (B), for example irradiation by the Sun, SO$_2$ can be abundant in the deep atmosphere, however in the upper atmosphere SO$_2$ is efficiently catalysed into H$_2$SO$_4$ by UV photons and subsequent reaction with H$_2$O \cite{YungDeMore1982}. In the Venus-like case with low-UV irradiation (C), such as irradiation by a low-UV M-dwarf, SO$_2$ can survive in the upper atmosphere \cite{Jordan2021}, potentially enabling SO$_2$ to be identified in spectroscopic observation and thereby revealing the lack of a surface water ocean. \label{fig:schematic}}
\end{figure}

In contrast, it has been demonstrated that the redder stellar spectra of M-dwarfs compared with Sun-like stars can be more amenable to the persistence of sulfur chemistry at low pressure in an exoplanet's atmosphere \cite{Jordan2021}. This potentially allows for the identification of surface uninhabitability on rocky exoplanets using observations with JWST. There are three different scenarios for the fate of SO$_2$ in rocky planet atmospheres (Fig.\,\ref{fig:schematic}): on wet temperate Earth-like planets, rainout and wet deposition sequesters SO$_2$ into surface oceans (Fig. \ref{fig:schematic}A); on dry Venus-like planets receiving high-UV fluxes photolysis of SO$_2$ sequesters sulfur into H$_2$SO$_4$ clouds preventing it from reaching the observable above-cloud atmosphere (Fig.\,\ref{fig:schematic}B);  however, on such dry planets with low-UV flux there is the possibility of inefficient photolysis and thus the persistence of an observable sulfur-cycle (the M-dwarf case, Fig.\,\ref{fig:schematic}C).  It is this last regime that we explore in detail in this paper.

The low-UV scenario offered by M-dwarfs does not explicitly allow us to identify Earth-like planets, where liquid water is present, however it does allow us to positively identify Venus-like planets, where liquid water is not present, via the detection of atmospheric SO$_2$. If there is an evolutionary transition between the Habitable Zone and the Venus Zone then we would expect that finding a Venus-like planet at an orbital distance $d$ from its host star negates the possibility of finding an Earth-like planet at some orbital distance $d' < d$ from the same host star. We therefore propose that mapping the presence of atmospheric SO$_2$ across a population of low-UV M-dwarf systems will provide an implicit test of the location of the inner edge of the Habitable Zone. While it has been demonstrated that the outer edge of the theoretically possible Venus Zone \cite{Vidaurri2022} overlaps with the inner edge of the theoretically possible Habitable Zone \cite{Kopparapu2013, Kopparapu2017, Leconte2013, WayDelGenio2020, Turbet2023}, the true \textit{empirical} transition between the Earth-like and Venus-like regimes will lie somewhere within this theoretical region of overlap, if the transition is a sharp boundary.

Previous work has investigated how population-level studies, utilising a comparative planetology approach, may be more effective at testing the model concepts underpinning the theory of the Habitable Zone, rather than the systems science approach to characterise individual targets in high detail \cite{Bean2017}. An extensively studied example is the expected trend of CO$_2$ partial pressure (pCO$_2$) with stellar irradiation within the Habitable Zone due to the carbonate-silicate weathering feedback \cite{Bean2017, Turbet2020, Checlair2019, GrahamPierrehumbert2020, Lehmer2020}. In this example, atmospheric pCO$_2$ on habitable planets should decrease with increasing irradiation up to the instellation limit of the inner edge of the Habitable Zone. Quantification of the CO$_2$ mixing ratio across a sample could therefore provide statistical evidence of a temperature stabilisation effect on Earth-like planets in the Habitable Zone. This approach requires a large enough sample of Earth-like planets so that uncertain parameters can be marginalised over, and requires constraining the abundance of CO$_2$, not just the presence of CO$_2$. Both requirements may be challenging with current observing facilities. An alternative approach for tracing the inner edge discontinuity is the radius inflation effect that occurs while planets are transitioning from wet and temperate to hot and dry, via an inflated steam atmosphere phase \cite{Turbet2019, Turbet2020, Schlecker2024}. This approach would provide an accurate estimate of the instellation limit at which runaway greenhouse is initiated, but requires catching planets in the act of transitioning, before the inflated steam atmosphere is lost by photodissociation and hydrogen-loss, as has occurred on Venus. As is the case for the uncertain parameters and sources of scatter in the pCO$_2$ trend, the timescale of the radius inflation effect is an uncertain parameter which must be marginalised over in any statistical inference, necessitating a sufficiently large sample size \cite{Turbet2019, Schlecker2024}.

These suggested comparative planetology studies, and other approaches which aim to constrain the habitability of individual targets such as detection of ocean glint \cite{LustigYaeger2018} or CO$_2$ dissolution in an ocean \cite{Triaud2023}, typically approach the problem from the perspective of the Earth-like, habitable regime, or a transition from the habitable regime. Here, we propose to focus instead on the opposing side of the problem --- the Venus-like, canonically uninhabitable regime. Identification of truly Earth-like exoplanets hosting liquid water oceans remains a fundamental goal of the community, however this is subject to observational degeneracies that will be difficult to resolve at the noise floor of JWST. The uninhabitable regime, in contrast, can be more directly linked to the observed planetary spectrum via the solubility equilibria of sulfur dioxide, provided that, crucially, the stellar radiation environment is suitable for the persistence of sulfur chemistry at observable pressure levels in a planet's atmosphere. In this paper, we explore the regions of stellar-UV parameter space where sulfur-chemistry can be used as an observational tracer of lack of oceans on rocky exoplanets orbiting M-dwarfs. We self-consistently simulate the climate and chemistry of Venus-like exoplanets (i.e., desiccated Venus-sized planets with high pressure CO$_2$-N$_2$ atmospheres, trace H$_2$O, and possible H$_2$SO$_4$ clouds) with varying sulfur-compositions, over a range of instellation fluxes that span all current estimates of the location of the inner edge of the liquid-water HZ \cite{Kopparapu2013,Kopparapu2017,Leconte2013,WayDelGenio2020} and WCZ \cite{Turbet2023}. For the stellar UV-distributions, we use the spectral data products from the MUSCLES and Mega-MUSCLES treasury surveys \cite{France2016, Youngblood2016, Loyd2016, Wilson2021}. Our results propose a roadmap toward mapping the inner edge of the \textit{empirical} Habitable Zone, thereby testing the various model concepts of habitability that have emerged throughout the literature.

\section*{Results}
\label{sec:results}

\subsection*{Stellar UV} 

The photochemical depletion of atmospheric gases by the irradiating starlight of an M-dwarf is controlled by two competing effects; the potentially enhanced extreme-UV (EUV) flux at short wavelengths, which increases photochemical activity, versus the lower stellar effective temperatures shifting of the peak stellar emission to longer wavelengths, diminishing photochemical activity. The net result of these competing effects leads to low-UV M-dwarfs allowing for the persistence of trace sulfur gases in an exoplanet's atmosphere, while, in contrast, high-UV M-dwarfs can preclude the existence of trace sulfur gases in the upper atmosphere of exoplanets, and instead form photochemical clouds and hazes that enhance planetary albedo and truncate observable spectroscopic features. In the case of the low-UV M-dwarf, the diagnosis of sulfur gases with spectroscopic observations may be possible despite observations only probing the upper atmospheric pressure levels.

\begin{figure}[b!]
    \centering
    \includegraphics[width=\textwidth]{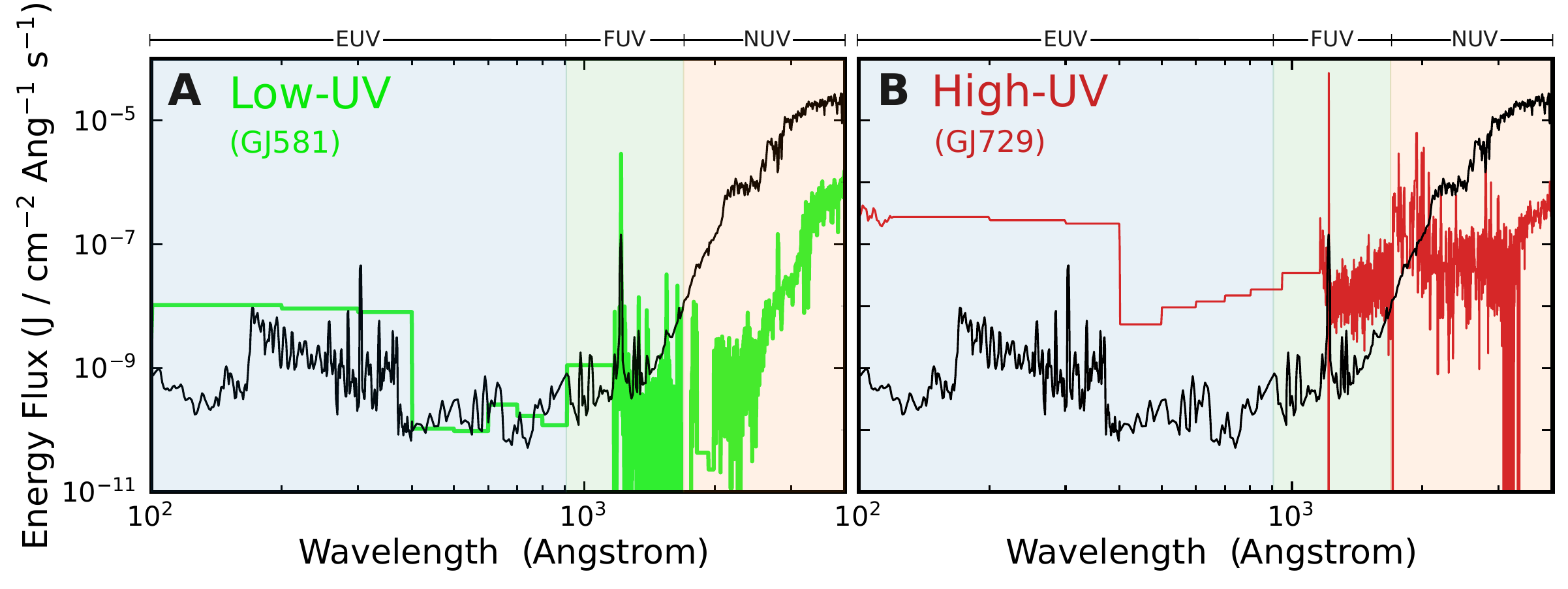}
    \caption{\textbf{Stellar UV distributions for a low-UV and a high-UV M-dwarf.} Stellar UV distribution of a low-UV M-dwarf (GJ581, T$_{\rm eff}$\,$\sim$\,3424\,K) (A) and a high-UV M-dwarf (GJ729, T$_{\rm eff}$\,$\sim$\,3248\,K) (B) obtained from the Mega-MUSCLES Treasury survey \cite{Wilson2021}. The Solar UV spectrum is plotted for reference in black. Each stellar spectrum is normalised to the bolometric luminosity that Venus receives from the Sun. \label{fig:spectra}}
\end{figure}

We demonstrate this effect here in Fig.\,\ref{fig:spectra} with two examples of M-dwarf stellar spectra from the Mega-MUSCLES survey: GJ581 (M3V) and GJ729 (M3.5V) \cite{Wilson2021}, each normalised to the bolometric luminosity that Venus receives from the Sun. The effective temperature of GJ581 is 3424\,K, which corresponds to a blackbody peak at 8463\,\AA. The EUV flux at shorter wavelengths is slightly higher than that of the Solar spectrum (Fig.\,\ref{fig:spectra}A, \textit{black}), however the red-wards shift of peak emission results in orders of magnitude lesser far-UV (FUV) and near-UV (NUV) flux irradiating a planet's atmosphere for the same bolometric luminosity. The resulting low-UV is not specific to GJ581: we highlight the GJ581 spectrum to represent the details of this general astrophysical effect. In fact, the majority of the M-dwarf stellar spectra captured in data products by the MUSCLES and Mega-MUSCLES surveys share a similar low-UV behaviour to that demonstrated here by GJ581. We return to this point and further generalise our results with respect to host star properties in the discussion section.

The shift in blackbody temperature to longer wavelengths does not always result in lower net UV emission, however, as demonstrated by the spectrum of GJ729 (Fig.\,\ref{fig:spectra}B) \cite{Wilson2021}. GJ729 has an effective temperature of 3248\,K, cooler than that of GJ581, which would result in a Wien peak at 8921\,\AA. The enhanced FUV and EUV flux of this star relative to its bolometric luminosity however, compensates for the reduction in UV flux that would otherwise be observed from the cooler blackbody-like emission. Using these spectra as inputs to our coupled climate-chemistry model, we now demonstrate how these different spectral inputs drive very different photochemistry in an exoplanet's atmosphere at instellation limits spanning model estimates of the HZ inner edge and the WCZ.

\subsection*{Photochemistry around the Runaway Greenhouse limit} 

\begin{figure}[t!]
    \centering
    \includegraphics[width=\textwidth]{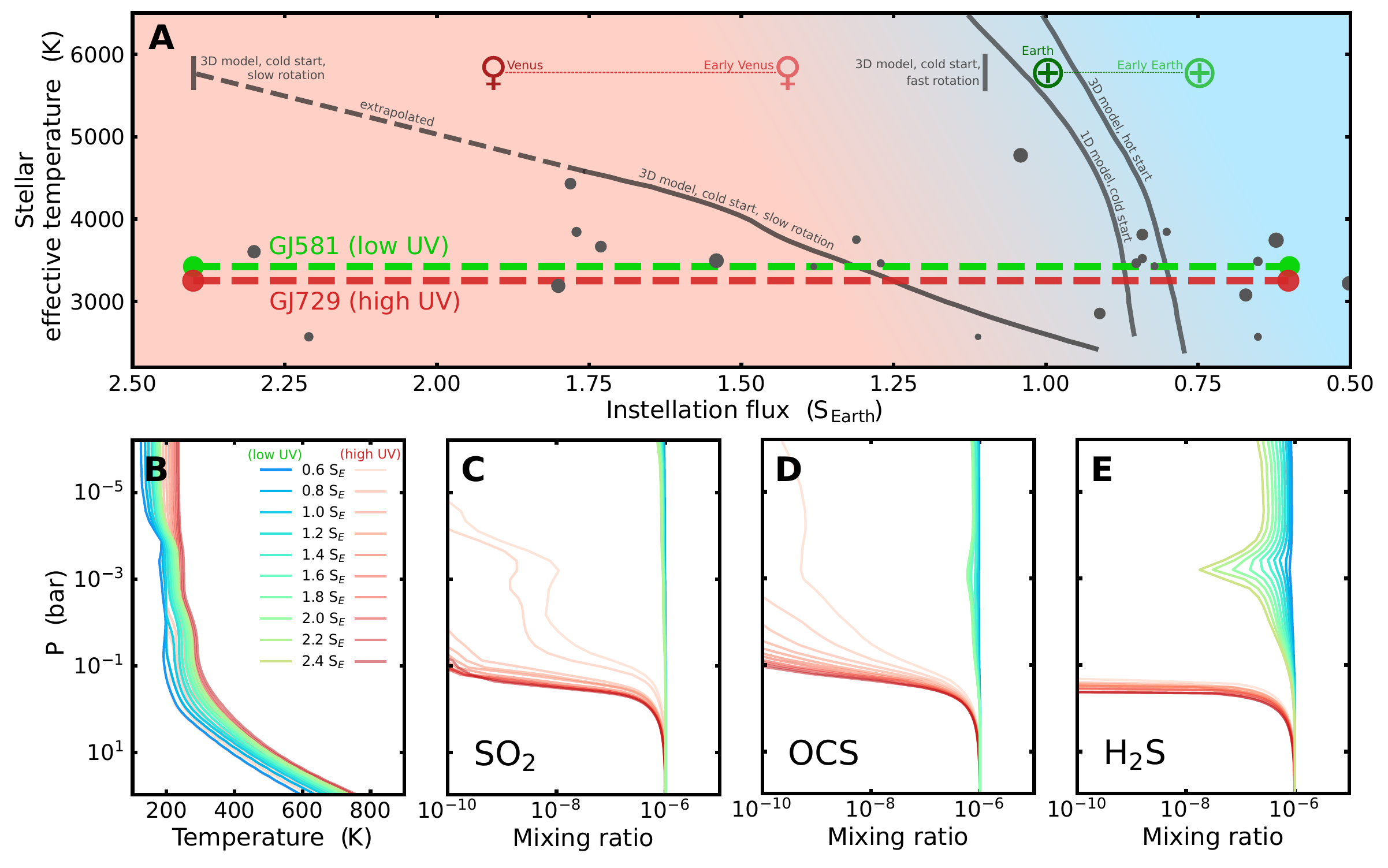}
    \caption{\textbf{Photochemistry around the runaway greenhouse limit for a low-UV and a high-UV M-dwarf.} (A) Estimated instellation limits of the inner edge of the liquid water Habitable Zone and Water Condensation Zone, as a function of stellar effective temperature of the host star. Model estimates, from left to right, correspond to: 3D model, cold start, slow rotation (Solar) \cite{WayDelGenio2020}; 3D model, cold start, slow rotation \cite{Kopparapu2017}; 3D model, cold start, fast rotation \cite{Leconte2013}; 1D model, cold start \cite{Kopparapu2013}; 3D model, hot start \cite{Turbet2023}. The instellation flux received by modern Venus, Early Venus, modern Earth, and Early Earth, are shown for reference, demonstrating our one pair of data points for the inner instellation limit of the \textit{empirical} Habitable Zone, at the Solar effective temperature T$_{\rm eff, \odot}$\,$\sim$\,5772\,K. Grey circles represent locations of exoplanets from the NASA exoplanet archive with radii\,$\leq$\,1.8\,R$_{\oplus}$, excluding systems further than 50\,pc. (B) Pressure-temperature profiles for a Venus-like atmosphere with 1\,ppm SO$_2$, irradiated by a low-UV M-dwarf (GJ581, blue\,--\,green) and a high-UV M-dwarf (GJ729, light red\,--\,dark red). Mixing ratios of SO$_2$ (C), OCS (D), and H$_2$S (E), as a function of atmospheric pressure, with the same colour scheme as panel B. \label{fig:depletion}}
\end{figure}

The shift in the peak blackbody emission of host stars with different stellar effective temperatures also influences the heating of planetary atmospheres and surfaces.  As a result, estimates of the HZ inner edge are a function of stellar effective temperature, moving outwards for cooler stars (Fig.\,\ref{fig:depletion}A). At a given stellar effective temperature, however, the instellation limit of the HZ inner edge varies according to modelling assumptions. The different model estimates of the inner edge therefore span a wide region of the stellar effective temperature versus instellation flux parameter space (Fig.\,\ref{fig:depletion}A). In this parameter space, Venus and Earth provide two concrete data points of a liquid-water uninhabitable planet and a habitable one respectively, at T$_{\rm eff, \odot}$\,$\sim$\,5772\,K. In order to constrain the location of the inner edge over the rest of the parameter space, we require observational tests of the climate and surface conditions on rocky worlds across the T$_{\rm eff}$ range. The preponderance of terrestrial objects discovered across the Venus Zone \cite{Kane2014,Ostberg2023} and short period regions of the HZ make this test possible in principle, provided that such objects have atmospheres and their surface conditions can be diagnosed with remote observation. We here demonstrate how the UV spectrum of M-dwarf host stars can enable this test across a population of suitably chosen targets via the photochemical behaviour of SO$_2$.

It has previously been shown that an atmosphere containing $\gtrsim$\,1\,ppm SO$_2$ requires $\lesssim$\,10$^{-3}$\,Earth ocean equivalent surface water in order to remain stable against wet deposition, using conservative assumptions about the oxidised sulfur-cycle on rocky planets \cite{Loftus2019}. The detection of SO$_2$ on rocky exoplanets is therefore a probe of uninhabitable surface conditions. Fig.\,\ref{fig:depletion} shows this case of 1\,ppm SO$_2$, for planets irradiated by the low-UV M-dwarf case (\textit{green}) compared to the high-UV M-dwarf case (\textit{red}). The corresponding tracks that these simulations probe in the stellar effective temperature versus instellation flux parameter space are shown with the green and red dashed lines in Fig.\,\ref{fig:depletion}A, based on the stellar effective temperatures of the low-UV and high-UV stars GJ581 and GJ729 respectively. In addition to SO$_2$ we also demonstrate how the other sulfur-containing gases, OCS and H$_2$S, respond to the irradiation of each stellar host. Like SO$_2$, the sulfur-species OCS and H$_2$S are photochemically sensitive to the UV variations between M-dwarfs, and are important sulfur gases on Venus, therefore they may also be potentially useful Venus-indicator gases on warm rocky exoplanets. We also show the resulting pressure-temperature profile alongside the atmospheric chemistry (Fig.\,\ref{fig:depletion}B). The pressure-temperature profile is important for resulting observables since it controls the altitude of cloud formation, thereby setting the baseline of spectroscopic observations, and it is correlated with the amplitude of spectral features by setting the atmospheric scale height.

Across the full instellation range we explore, each sulfur-gas, SO$_2$ (Fig.\,\ref{fig:depletion}C), OCS (Fig.\,\ref{fig:depletion}D), and H$_2$S (Fig.\,\ref{fig:depletion}E), survives to observable pressure levels when irradiated by the low-UV M-dwarf. In contrast, each sulfur-gas is efficiently photochemically destroyed above $\sim$\,0.1\,--\,1\,bar when irradiated by the high-UV M-dwarf. This divergence in upper atmosphere chemistry, resulting from a change in only the UV distribution at a given bolometric flux, evidences how it is imperative that stellar UV is well characterised for both predicting observables and explaining observed atmospheric features \textit{post-hoc}. In the context of mapping the inner edge of the Habitable Zone, Fig.\,\ref{fig:depletion} reveals that, for $\sim$\,1\,ppm SO$_2$, this would be possible in principle for the GJ581 case of low-UV, and would not be possible for the GJ729 case of high-UV. We now investigate how these differences translate into observables that can be accessed by JWST.

\subsection*{Observational signals of sulfur-chemistry} 

\begin{figure}[ht!]
    \centering
    \includegraphics[width=\textwidth]{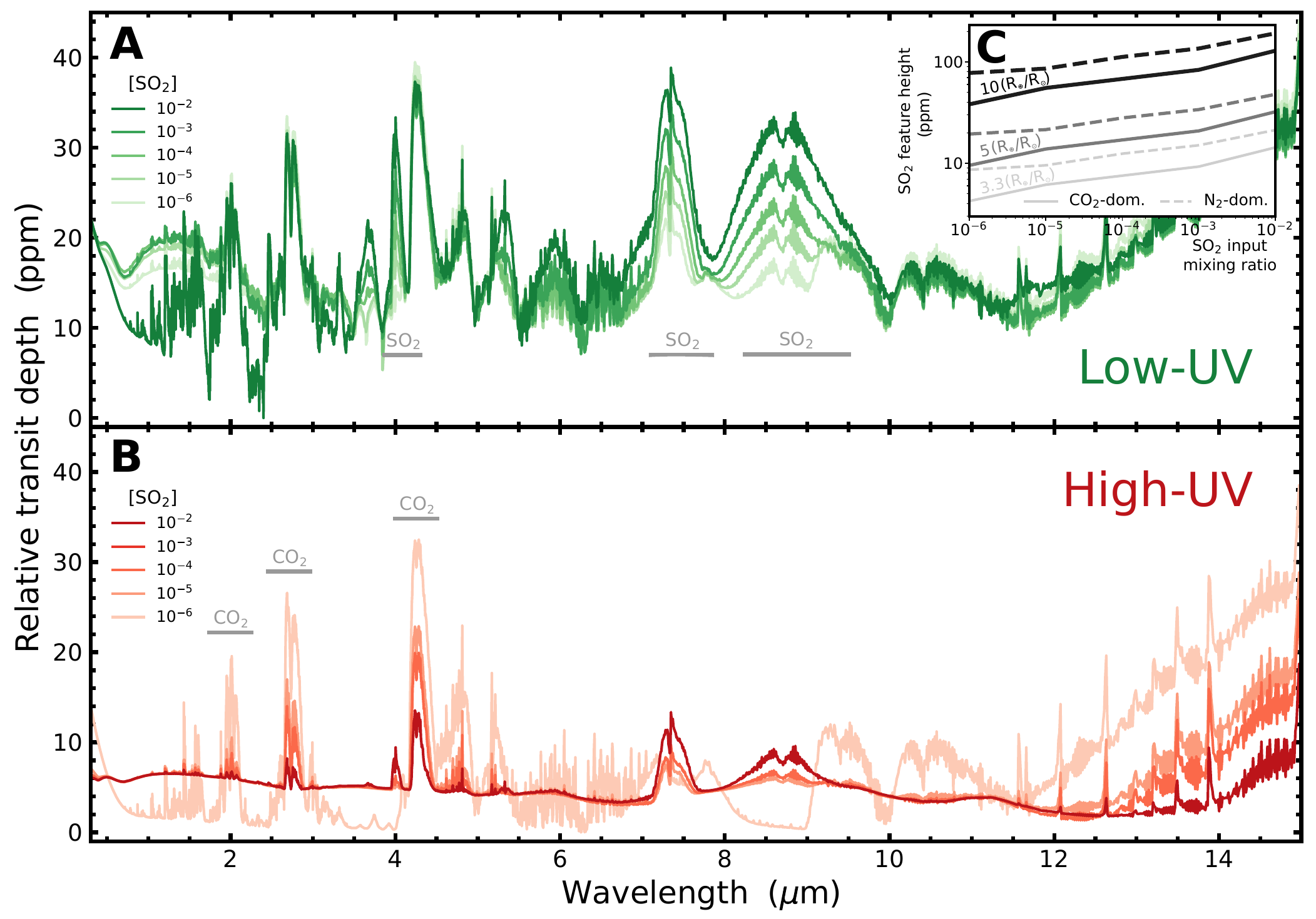}
    \caption{\textbf{Observable signals of sulfur chemistry.} Transmission spectra for Venus-like atmospheres with varying SO$_2$ abundance, at an instellation flux of 2\,S$_{\oplus}$. The atmospheres are irradiated with low-UV flux (A) and high-UV flux (B). Transmission spectra are simulated using petitRADTRANS \cite{Molliere2019,Molliere2020,Alei2022}, where transit depth is calculated using a planet-star radius ratio of (R$_{\oplus}$/0.2\,R$_{\odot}$\,=\,5\,R$_{\oplus}$/\,R$_{\odot}$). Wavelength ranges with prominent SO$_2$ features (A) and CO$_2$ features (B) are indicated in grey. The inset (C) shows how the maximum transmission feature height due to SO$_2$ varies for a CO$_2$-dominated versus N$_2$-dominated background atmosphere, and for three different planet-star size ratios (10\,R$_{\oplus}$/R$_{\odot}$, 5\,R$_{\oplus}$/R$_{\odot}$, and 3.3\,R$_{\oplus}$/R$_{\odot}$). The CO$_2$-dominated background atmosphere is composed of $\sim$\,(96.5\% CO$_2$, 3.5\% N$_2$), and the N$_2$-dominated background atmosphere is composed of $\sim$\,(3.5\% CO$_2$, 96.5\% N$_2$). \label{fig:transit_spectra}} 
\end{figure}

The observability of transmission features due to SO$_2$ in a Venus-like exoplanet's atmosphere depends crucially on the host star UV flux (Fig.\,\ref{fig:transit_spectra}). Transmission features due to SO$_2$ are prominent at 4\,$\mu$m, 7\,--\,8\,$\mu$m, and 8\,--\,9.5\,$\mu$m, while transmission features due to CO$_2$ are prominent at 2\,$\mu$m, 2.8\,$\mu$m, and, most strongly, at 4.3\,$\mu$m. In the low-UV case, the height of the 7\,--\,8\,$\mu$m SO$_2$ feature approaches the height of the prominent 4.3\,$\mu$m CO$_2$ feature (Fig.\,\ref{fig:transit_spectra}A). At 1\,ppm SO$_2$ in the deep atmosphere, the SO$_2$ feature is one third of the height of the CO$_2$ feature; at 100\,ppm SO$_2$ in the deep atmosphere, it is over half the height of the CO$_2$ feature; by 1\% atmospheric SO$_2$, the 7\,--\,8\,$\mu$m SO$_2$ feature is the tallest peak in the spectrum (Fig.\,\ref{fig:transit_spectra}A).

The high-UV case, however, tells a very different story (Fig.\,\ref{fig:transit_spectra}B). Whilst the height of SO$_2$ features increases with increasing abundance of trace SO$_2$ in the atmosphere, the continuum spectral baseline also increases. This increase in both spectral features and baseline ultimately leads to smaller changes in the height of the SO$_2$ spectral features compared to the low-UV case. This is because the high-UV flux catalyses the conversion of some or all of the atmospheric SO$_2$ into condensible H$_2$SO$_4$. The H$_2$SO$_4$ clouds that then form in the atmosphere truncate the transmission spectrum and cancel out gains in the height of SO$_2$ features with increasing abundance in the deep atmosphere. This cloud truncation can also be seen in the 4.3\,$\mu$m CO$_2$ feature: as atmospheric SO$_2$ increases, the relative height of this CO$_2$ feature to the cloud continuum decreases (Fig.\,\ref{fig:transit_spectra}B). The SO$_2$ feature grows slowly while the CO$_2$ feature diminishes quickly until, by $\sim$\,1\% atmospheric SO$_2$, the above-cloud SO$_2$ and CO$_2$ spectral features become closely comparable in height, and both far smaller than in the low-UV case.

These results evidence the crucial differences that emerge for observable sulfur-chemistry on exoplanets orbiting M-dwarfs generally. The possibility for SO$_2$ features to remain observable in a dry exoplanet's atmosphere requires low-UV due to photochemical conditioning of the upper atmosphere, however the depth of the resulting spectroscopic features will also depend on the atmospheric scale height and planet-star size ratio. The case shown in Fig.\,\ref{fig:transit_spectra} assumes a planet-star ratio of 5\,R$_{\oplus}$/\,R$_{\odot}$, equivalent to an Earth-sized planet around a 0.2\,R$_{\odot}$ star, or a 1.5\,R$_{\oplus}$ super-Earth around a 0.3\,R$_{\odot}$ star. We further explore how the transit depth of the tallest SO$_2$ feature in the transmission spectrum varies with planet-star size ratio and CO$_2$- vs N$_2$-dominated background atmospheres in Fig.\,\ref{fig:transit_spectra}C. For a planet-star size ratio of 10\,R$_{\oplus}$/\,R$_{\odot}$ the SO$_2$ feature height reaches $\gtrsim$\,100\,ppm  in transit depth when the deep atmosphere SO$_2$ mixing ratio is $\gtrsim$\,2000\,ppm in a CO$_2$-dominated background, or $\gtrsim$\,30\,ppm in an N$_2$-dominated background. We conclude that low-UV M-dwarf hosts with stellar radii $\lesssim$\,0.3\,R$_{\odot}$ will enable transmission observations of SO$_2$ features on the order of 10s\,--\,100s\,ppm, accessible with order 10s of transits captured by JWST \cite{LustigYaeger2023}. This estimate of observability assumes a reasonable sulfur abundance in these planets' atmospheres. Although sulfur is moderately volatile, it is abundant at the surface environments of all rocky planets, from Mercury to Mars, suggesting it is common in rocky planet building blocks \cite{kama2019abundant}.

This result enhances the `M-dwarf opportunity', now defined explicitly with respect to the atmospheric chemistry of their planets, opening up the opportunity to constrain exoplanet surface conditions and map the HZ inner edge via sulfur-chemistry [e.g., \cite{Loftus2019, Byrne2024}]. This is otherwise not possible for exoplanets around Solar-type stars and thus has often not been considered in investigations of Venus-like exoplanets. Motivated by this, we now investigate the detectability of exoplanet sulfur-chemistry in detail over a wide parameter space of instellation flux and sulfur abundance.

\subsection*{Cloud albedo and transit features across instellation space} 

\begin{figure}[htbp!]
    \centering
    \includegraphics[width=0.9\textwidth]{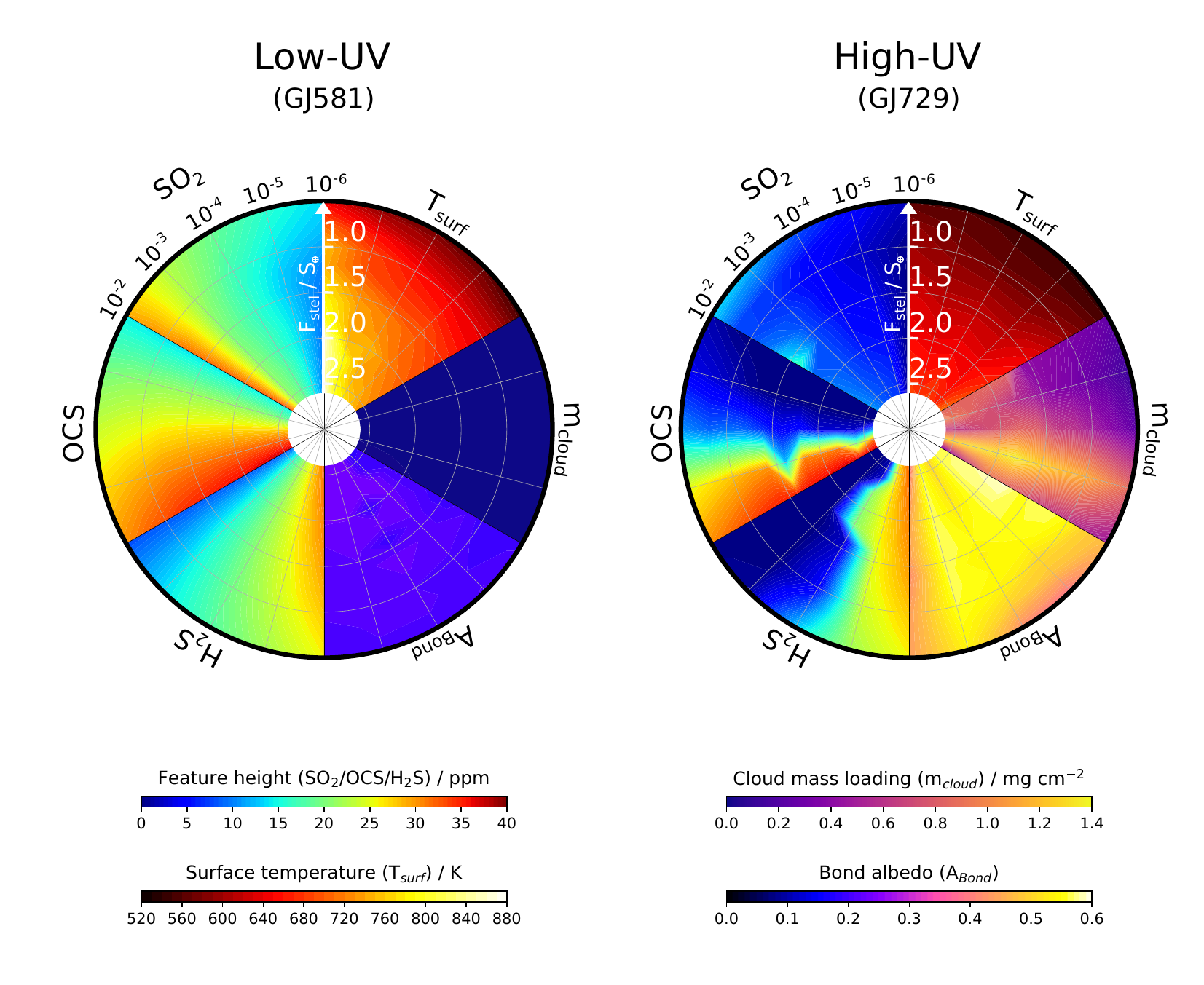}
    \caption{\textbf{Transit features and cloud albedo for different atmospheric sulfur abundances.} Polar plots showing surface temperature, cloud mass, Bond albedo, and the transmission feature heights of the sulfur-gases SO$_2$, OCS, and H$_2$S, in a Venus-like atmosphere irradiated by a low-UV M-dwarf and a high-UV M-dwarf. From the top, going clockwise, segments of the polar plots show: surface temperature, T$_{\rm surf}$\,(K); cloud mass loading, m$_{\rm cloud}$\,(mg\,cm$^{-2}$); Bond albedo, A$_{\rm Bond}$; H$_2$S transmission feature height\,(ppm); OCS transmission feature height\,(ppm); SO$_2$ transmission feature height\,(ppm). The radial coordinate corresponds to decreasing instellation flux (or increasing distance from host star) in units S$_{\oplus}$. The azimuthal coordinate, going anti-clockwise around the edge of each segment, corresponds to increasing mixing ratio of the input sulfur-species, from 10$^{-6}$ (1\,ppm) up to 10$^{-2}$ (1\%). The T$_{\rm surf}$, m$_{\rm cloud}$, and A$_{\rm Bond}$ segments are plotted for the case where the azimuthal coordinate is tracking an increase of the input SO$_2$ abundance anti-clockwise.  For the SO$_2$, OCS, and H$_2$S feature height segments, the azimuthal coordinate corresponds to the input abundance of SO$_2$, OCS, and H$_2$S, respectively, with the other sulfur-gases not being input to the atmosphere. \label{fig:polar}}
\end{figure}

We have demonstrated how the height of the 7\,--\,8\,$\mu$m SO$_2$ feature in transmission spectroscopy will be a particularly accessible probe of atmospheric SO$_2$, and thus lack of water oceans, in M-dwarf systems where the UV flux does not efficiently catalyse the conversion of SO$_2$ into H$_2$SO$_4$ (Fig.\,\ref{fig:transit_spectra}A). In systems where the host star UV flux does efficiently catalyse this conversion, the amount and altitude of the resulting cloud formation has a strong influence in flattening the transit spectrum (Fig.\,\ref{fig:transit_spectra}B). We now investigate these effects over the full parameter space of instellation flux and input sulfur-species abundance for each of SO$_2$, OCS, and H$_2$S, and contrast the results between low-UV case and high-UV case  (Fig.\,\ref{fig:polar}).

The primary observational point demonstrated in Fig.\,\ref{fig:polar} is that the sulfur gases are all observable in the atmosphere of the low-UV case, with the height of a given spectral feature correlating primarily with the abundance of the sulfur species in the deep atmosphere, and secondarily with the atmospheric temperature. In particular, the feature height for OCS is generally the strongest feature in the transmission spectrum and may be a powerful observational probe of sulfur chemistry on exoplanets. The high-UV case, in contrast, efficiently photochemically depletes all three sulfur gases when their atmospheric mixing ratios are $\lesssim$\,1000\,ppm in the deep atmosphere, such that they will not be spectroscopically accessible.

We note that there is an added subtlety to the high-UV case: for the SO$_2$, OCS, and H$_2$S segments, these gases are each input in separate models so that their individual thermochemical and photochemical behaviour is reflected in the observability, rather than inputting the three gases at once. Fig.\,\ref{fig:polar} reveals that the SO$_2$ feature heights with high-UV irradiation remain minimal across all the input abundances whereas the OCS and H$_2$S feature heights are observable for atmospheric abundances greater than $\gtrsim$\,1000\,ppm. This is due to the chemistry of the cloud formation. SO$_2$ is the precursor to H$_2$SO$_4$ production, and when H$_2$SO$_4$ cloud layers form, they truncate the observable region of the atmosphere (Fig.\,\ref{fig:transit_spectra}B). Any SO$_2$ which remains above the clouds due to self-shielding then only results in muted transmission features. In contrast, OCS and H$_2$S are not direct cloud precursors so this atmospheric truncation in transmission spectroscopy does not occur. If SO$_2$ was being input to the atmosphere at the same time as the OCS and H$_2$S, even at lower abundance, this effect may still occur and truncate the observable atmosphere with photochemical cloud formation. It is unlikely that the three sulfur gases would be input to an atmosphere in equal abundance from volcanic degassing, however it is likely that they will be degassed simultaneously at some ratio determined by the interior mantle chemistry [e.g., \cite{liggins2022}].

The second point demonstrated in Fig.\,\ref{fig:polar} that is critical to observations is the difference in cloud mass loading and resultant Bond albedo between the two M-dwarf cases. The low-UV case has very little cloud formation, resulting in a Bond albedo of $\sim$\,0.15\,--\,0.25 across the whole parameter space. Conversely, in the high-UV case the cloud formation and Bond albedo is very different. The cloud mass loading is significant at all input SO$_2$ abundances which results in Bond albedos ranging from $\sim$\,0.45\,--\,0.60. Whilst the transit feature heights for SO$_2$ in the high-UV case may be inaccessible in primary eclipse with JWST, the Bond albedo is starkly different to that expected from a cloud-free CO$_2$ atmosphere, and could be constrained in secondary eclipse. This means that the divergent atmospheric chemistries shaping the observable regions of the atmosphere around high-UV vs low-UV M-dwarf host stars do not necessarily preclude sulfur-chemistry being used to map the space of uninhabitable planets. Instead, these results emphasise that different observing methods should be used to draw the habitable/uninhabitable distinction. Degeneracies nonetheless arise in the high Bond albedo case, and we return to this point further in the discussion section. Our results here, in any case, demonstrate the rich space of sulfur chemistry that can be explored observationally on rocky worlds with JWST.

\section*{Discussion}
\label{sec:discussion}

Our results demonstrate how the observability of sulfur gases and hazes depend crucially on the UV spectrum of M-dwarf host stars, a property which shows a wide variation. In particular, we highlight the unintuitive result that an M-dwarf's stellar radiation can be more amenable for the persistence of sulfur-gases high in an exoplanet's atmosphere than the spectrum of higher effective temperature stars, such as the Sun. The sulfur species that we have demonstrated this for are SO$_2$, OCS, H$_2$S, and condensible H$_2$SO$_4$. Below we discuss the significance of observing each of these species in an exoplanet's atmosphere and the link that may be drawn between these astronomical observations to inferences about surface habitability.

\subsection*{Observing SO$_2$ with spectroscopy} 

The presence of SO$_2$ in an oxidised exoplanet's atmosphere has been previously linked to the lack of substantial surface liquid water \cite{Loftus2019}. Our results have demonstrated that the presence of SO$_2$ features in the transmission spectra of oxidised exoplanets, and thus an empirical diagnosis of lack of surface water, will be possible to observe in systems with low-UV host stars. High-UV host stars, in contrast, eradicate trace SO$_2$ at pressure levels in the atmosphere where transmission spectroscopy probes, and lead to H$_2$SO$_4$ cloud formation which truncates spectroscopic features. Exoplanets orbiting M-dwarf host stars can have their atmospheres resolved with $\sim$\,10\,s of transits captured with JWST and their planetary evolution is expected to yield oxidised atmospheres and interiors \cite{Wordsworth2018}. Our results reveal that M-dwarfs which do not have an enhanced UV flux are photochemically suitable to enable SO$_2$ to be observed in the atmospheres of orbiting exoplanets, due to the red-wards shift of the peak stellar emission. We therefore propose that population-level spectroscopic searches aiming to map the bounds of surface uninhabitability on M-dwarf planets aim to constrain SO$_2$ mixing ratios on targets where the stellar UV is low. This recommendation may also extend beyond the sulfur species that we have examined in detail here to non-sulfur species useful for constraining the Habitable Zone, such as CO$_2$ and H$_2$O \cite{Bean2017,GrahamPierrehumbert2020}.

Current observing programs for rocky exoplanets are focused on answering whether or not rocky planets orbiting M-dwarfs can retain substantial atmospheres on short-period orbits. Planets on short-period orbits can have their atmospheres identified with limited JWST observing time because the planets transit more frequently and the planetary emission flux is more likely to be high. The observations performed so far have confirmed that short-period rocky planets do not retain substantial low mean-molecular weight atmospheres, but may be consistent with high mean molecular weight atmospheres and cloud-decks \cite{LustigYaeger2023,zieba2023,Moran2023,May2023,Kirk2024}. As more data are obtained and the prevalence of high mean molecular weight atmospheres on rocky exoplanets becomes better understood, the next frontier will be constraining exoplanet climates and surface conditions. This frontier will require pushing JWST observing capabilities to further distances from the host star where transmission spectroscopy may be more favourable than emission spectroscopy \cite{LustigYaeger2019}. In this observing regime, seeking transmission features of SO$_2$ has a number of advantages: SO$_2$ features can be sought with JWST's NIRSpec/PRISM, NIRSpec/G395H, and MIRI/LRS, demonstrated already by observations of WASP-39b \cite{Alderson2023, Tsai2023, Savvas2023, Powell2024}; SO$_2$ is not likely to be mimicked by star spots, unlike H$_2$O which can form in cool M-dwarf atmospheres and requires observing a broad wavelength range to rule out \cite{Kirk2024}; SO$_2$ transmission features can be of comparable transit depth as the CO$_2$ features from a Venus-like background atmosphere at only trace atmospheric abundances of $\sim$\,100s\,--\,1000s\,ppm. Constraining SO$_2$ and CO$_2$ in conjunction may be a powerful probe of whether a rocky exoplanet has undergone a Venus-like evolution.  Such observations would also be capable of ruling out the possible habitable climate state of a short-period slow-rotator [Fig.\,\ref{fig:depletion}; \cite{Yang2013, WayDelGenio2020}].

\subsection*{Observational implications of H$_2$SO$_4$ clouds} 

The prediction of observable SO$_2$ features in an M-dwarf exoplanet's atmosphere is not guaranteed, but depends crucially on detailed photochemical modelling of the planet-star system. For host stars with enhanced relative UV flux, strong photochemical depletion of SO$_2$ can lead to formation of H$_2$SO$_4$ cloud/haze layers. If this is the case then spectroscopic evidence of SO$_2$ present in the atmosphere can not be easily diagnosed even when the SO$_2$ abundance  maintained in the deeper atmosphere reaches up to 1\%\,: at low SO$_2$ abundances, all of the gaseous SO$_2$ is depleted in the observable region of the atmosphere; at high SO$_2$ abundances, formation of H$_2$SO$_4$ clouds at high altitudes truncates how much of the atmosphere is observable at all, in spite of any remaining SO$_2$ that may prevail from self-shielding.

While heavy photochemical depletion of SO$_2$ precludes its detection via spectroscopy, its photochemical product H$_2$SO$_4$ can potentially be inferred instead from the measured dayside albedo in secondary eclipse photometry \cite{Mansfield2019}. Photometric observations in secondary eclipse are far less time-consuming to obtain with JWST compared to obtaining atmospheric spectra, and can be used to infer the brightness temperature and thus Bond albedo of a planet \cite{Mansfield2019}. The Bond albedos that we obtain for a hypothetical Venus-like planet orbiting a low-UV M-dwarf versus a high-UV M-dwarf are around $\sim$\,0.15\,--\,0.25 and $\sim$\,0.45\,--\,0.60 respectively across the compositional range we have investigated (Fig.\,\ref{fig:polar}). Constraining Bond albedo with secondary eclipse photometry is therefore a readily available observation that can be made of rocky exoplanets orbiting M-dwarfs stars that have enhanced UV.

Diagnosing the lack of surface water oceans from a high H$_2$SO$_4$ cloud albedo is, however, more ambiguous than constraining SO$_2$ spectroscopically. The primary concern with inferring H$_2$SO$_4$ clouds from high albedo is that water clouds and reflective planetary surfaces can also lead to an observed low brightness temperature and therefore inference of high albedo \cite{Mansfield2019}. To observationally constrain surface habitability and the onset of runaway greenhouse, we must be able to disentangle planets that have passed through the runaway greenhouse boundary, or formed inside the water-condensation limit, versus planets that remain habitable inside the Venus Zone. Habitable Venus-zone planets can have their climates stabilised by highly reflective substellar water clouds \cite{Yang2013,WayDelGenio2020}. Additionally, glaciated planets with reflective water-ice at the surface and thin Earth-like, Mars-like, or even more tenuous atmospheres, may stabilise cold surface conditions at high instellation flux \cite{Konrad2023,Wordsworth2021Snowball}. One complimentary approach for disentangling these possible scenarios is to put observations of planetary albedo in context of climate modelling to theoretically rule out unstable cases. Another complimentary method may be the combination of inferred albedo from photometry with chemical information in spectroscopy that could build up a multi-phase picture of the atmospheric sulfur-cycle. This nonetheless may require next-generation instruments such as the Large Interferometer for Exoplanets (LIFE) \cite{Quanz2022} or the Habitable Worlds Observatory (HWO) \cite{Gaudi2020} due to the generally small spectral features observable against the high-altitude cloud continuum.

\subsection*{Accurate characterisation of stellar UV} 

The influence that the photochemistry of sulfur has on observations of planetary atmospheres points to the crucial importance of accurately determining the stellar energy distribution (SED) of planet-hosting stars. This point is well demonstrated by the example of the Trappist-1 system. Trappist-1 is a small, ultracool M-dwarf star, with an effective temperature of 2619\,K, a stellar radius of only 0.117\,R$_{\odot}$, and seven transiting terrestrial exoplanets \cite{Gillon2016}. The particularly small size of Trappist-1 leads to a high planet-star size ratio which aids atmospheric observations of the Trappist-1 planets. Because of the observational opportunity that the Trappist-1 system presents it has been the focus of many previous modelling and observational studies [e.g., \cite{Lincowski2023,Greene2023,zieba2023,BennekeTrappist-1g}]. This has necessitated the characterisation of Trappist-1's SED, and different empirical and semi-empirical data products have been presented so far in the literature \cite{Peacock2019,Wilson2021}.

Obtaining accurate observational constraints on the UV spectrum of Trappist-1 has proved difficult however due to its intrinsically faint UV emission as observed from the Earth \cite{Wilson2021}.  The low luminosity of Trappist-1 results in low signal-to-noise observations of its SED, particularly in the NUV \cite{Wilson2021}. To account for this observational limitation, Wilson et al., \cite{Wilson2021} produced a semi-empirical model of the Trappist-1 SED replacing the low signal-to-noise NUV flux with a model NUV distribution, for use in planetary atmosphere codes. UV observations of Trappist-1 taken since with HST will soon elucidate the true NUV flux of Trappist-1 \cite{WilsonHST}. We now demonstrate the importance of such observational stellar-UV constraints for the Trappist-1 system, and as a representative case study more generally.

\begin{figure}[htbp!]
    \centering
    \includegraphics[width=0.9\textwidth]{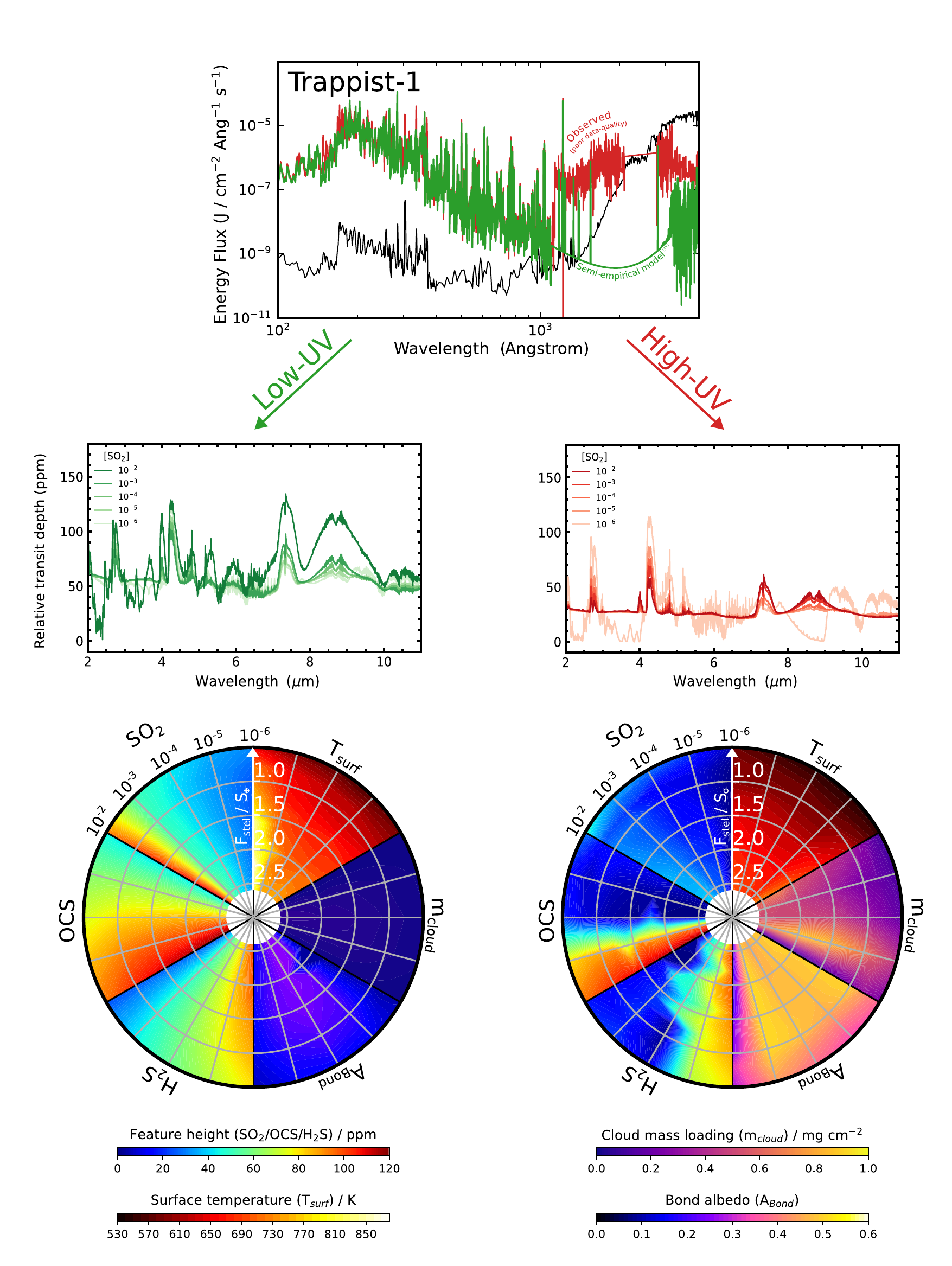}
    \caption{\textbf{Observational signals of sulfur chemistry for a planet orbiting Trappist-1 depends on the stellar NUV flux.} Repeat of our analysis highlighting how the low-UV and high-UV estimates of the Trappist-1 SED lead to different atmospheric sulfur chemistry and potentially divergent observables. Calculations of transit depth are done assuming a 1\,R$_{\oplus}$ planet around a 0.117\,R$_{\odot}$ star. \label{fig:trappist}}
\end{figure}

The observed SED and the semi-empirical SED for the UV spectrum of Trappist-1 propagate into divergent atmospheric sulfur-chemistry in an exoplanet's atmosphere (Fig.\,\ref{fig:trappist}). The observed NUV spectrum (\textit{red}) and the semi-empirical model (\textit{green}) have order-of-magnitude differences between the wavelength range $\sim$\,1200\,--\,4000\,\AA. This difference alone results in a divergence between the observable features of SO$_2$ and CO$_2$ in transmission spectroscopy due to photochemistry. For the low-UV case features due to SO$_2$, OCS, and H$_2$S are observable and reach up to $\gtrsim$\,100\,ppm in transit depth in a CO$_2$-dominated background atmosphere, due to Trappist-1's particularly small size. In the high-UV case the features from SO$_2$ remain muted to $\lesssim$\,40\,ppm regardless of the input SO$_2$ abundances due to the formation of clouds. The formation of H$_2$SO$_4$ clouds truncates spectroscopic observations that could be obtained from the Trappist-1 planets, but raises the Bond albedo to $\sim$\,0.40\,--\,0.55.

Our analysis reveals that different observational methods will be required to constrain the surface habitability of the Trappist-1 planets depending on whether the NUV flux of Trappist-1 is confirmed to be high or low. If Trappist-1 has a high NUV flux and any of the Trappist-1 planets are found to have a high albedo inferred from secondary eclipse \cite{Mansfield2019} then spectroscopic follow-up to constrain the possible presence or lack of surface water must be able to resolve above-cloud SO$_2$ features $\lesssim$\,40\,ppm in amplitude. Alternatively, if Trappist-1 is confirmed to have a low-NUV flux then the Trappist-1 planets are unlikely to form substantial H$_2$SO$_4$ clouds: if a high albedo is inferred in secondary eclipse then this could be due to a highly reflective surface ice and thin atmosphere, or stabilising substellar water clouds over a habitable surface; if a low albedo but presence of an atmosphere are inferred then the atmosphere will be favourable for characterising sulfur gases. In this case, the presence of atmospheric SO$_2$ would then unambiguously indicate a lack of surface water. The absence of SO$_2$ could possibly be linked to the presence of surface water however this would be a more ambiguous conclusion because the lack of SO$_2$ is not exclusively linked to wet deposition. For example, rocky planets could lack observable atmospheric SO$_2$ due to an alternative sink of sulfur gas in the atmosphere, or a very sulfur-poor formation scenario.

The observational implications of the dichotomous photochemistry of sulfur gases on the Trappist-1 planets highlights the importance of characterising the host star SED in all planetary systems. Data products such as the MUSCLES spectra are frequently employed as proxies for other host stars that have not been well characterised but share similar stellar parameters, such as age or effective temperature. The intrinsic variation of M-dwarf UV flux between the set of MUSCLES stars therefore highlights not only the importance of the characterisation of host star SED, but also the choice of proxy SED whenever observations of the specific host star are unavailable. Ongoing characterisation of the host stars in JWST target systems will be a crucial component of the era of atmospheric characterisation of small planets that we have now entered.

\subsection*{A population-level test of habitability} 

Current observations of rocky planets being made with JWST are aiming to answer whether rocky exoplanets on short period orbits ($\gtrsim$\,10\,S$_{\oplus}$ [e.g., \cite{hotrocks}]) can hold onto substantial atmospheres or not. The next frontier for rocky exoplanet characterisation will be longer period planets, which are more likely to retain secondary atmospheres, but require more observation time to characterise. For rocky exoplanets that have retained secondary atmospheres, our results have demonstrated how UV-driven photochemistry has a strong influence on the observable sulfur chemistry of M-dwarf exoplanets. Here we generalise our results beyond the examples demonstrated by GJ581, GJ729, and Trappist-1, to the full sample of MUSCLES M-dwarf and K-dwarf SEDs \cite{France2016, Youngblood2016, Loyd2016, Wilson2021}.

\begin{figure}[ht!]
    \centering
    \includegraphics[width=0.8\textwidth]{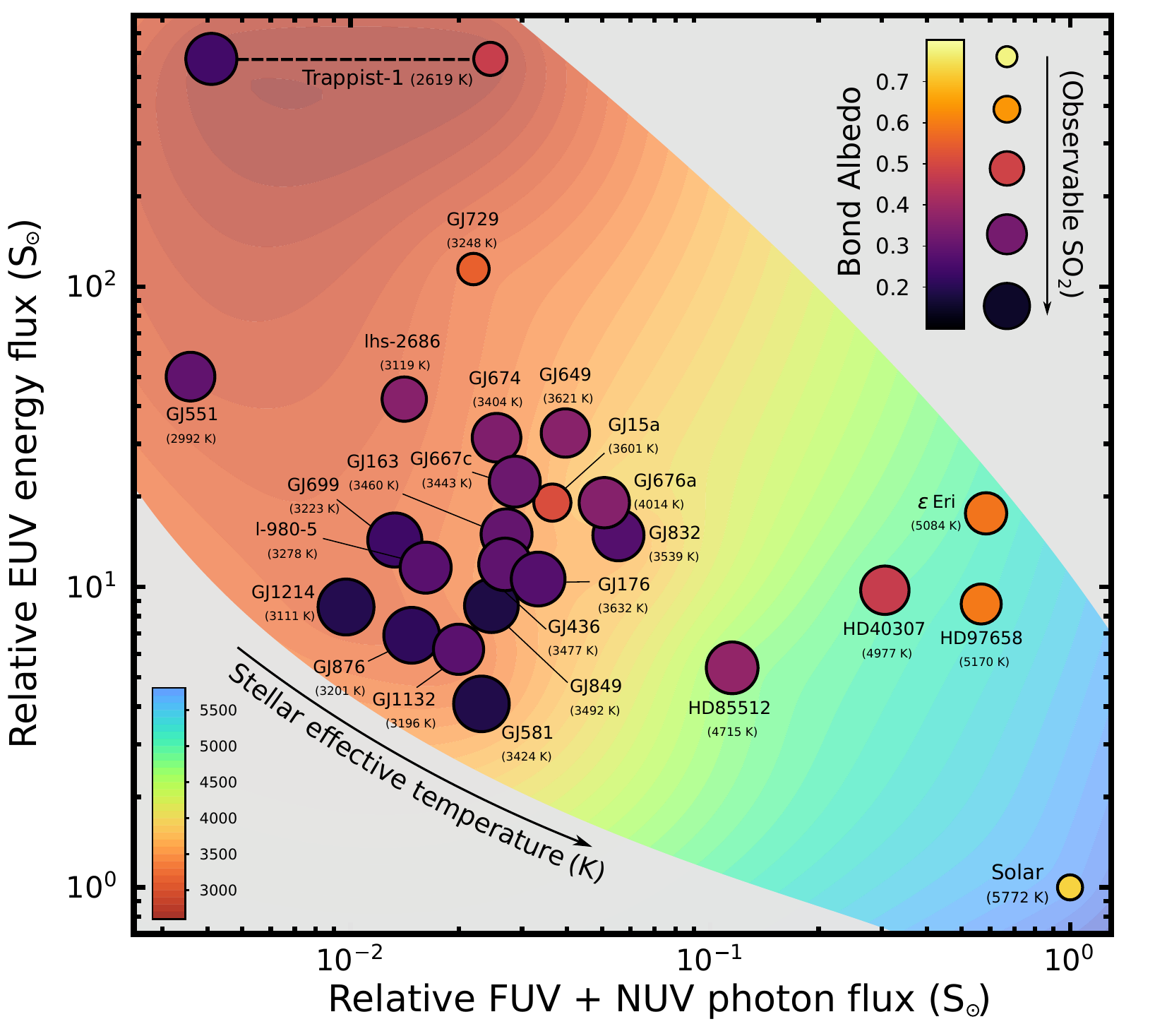}
    \caption{\textbf{UV fluxes of well characterised host stars, and their influence on observing sulfur cycles on exoplanets.} EUV flux versus FUV \& NUV flux of stellar spectra from the MUSCLES and Mega-MUSCLES Treasury surveys \cite{France2016,Youngblood2016,Loyd2016,Wilson2021}, relative to the Solar spectrum. Points are colour-coded by the Bond albedo that a Venus-like planet with 100\,ppm SO$_2$ in the atmosphere would have at an instellation flux of 1.6\,S$_{\oplus}$. The relative size of the circles shows qualitatively how the SO$_2$ transmission features vary with the different UV fluxes, for a fixed planet-star size ratio. The background colour shows a 2D interpolation of the stellar effective temperatures, contoured at intervals of 100\,K; within the data domain the points follow a clear trend of low stellar effective temperature in the top left corner, to high stellar effective temperature in the bottom right corner. \label{fig:star_plot}}
\end{figure}

The MUSCLES SED's demonstrate a wide range of EUV, FUV and NUV fluxes (Fig.\,\ref{fig:star_plot}). The EUV energy flux of the star is linked to escape processes for secondary outgassed atmospheres \cite{Owen2019}, and the FUV\,\&\,NUV photon flux is linked to the photochemistry of sulfur gases. Most of the M-dwarfs in the MUSCLES sample cluster around an FUV\,\&\,NUV flux 1\,--\,2 orders of magnitude lower than the that of the Sun. This results in a majority of the MUSCLES M-dwarf SEDs being favourable for the observation of sulfur gases spectroscopically in the atmospheres of hosted exoplanets. The SEDs that lead to favourable characterisation of sulfur gases also result in low Bond albedos between $\sim$\,0.1\,--\,0.3, due to inefficient formation of H$_2$SO$_4$ clouds. A number of outlying M-dwarf SEDs puncture this general trend however, namely GJ15a, GJ729, and the high-UV Trappist-1 estimate, all of which result in higher Bond albedos, between $\sim$\,0.4\,--\,0.6, and unfavourable conditions for characterising sulfur gases spectroscopically in an exoplanet's atmosphere (Fig.\,\ref{fig:star_plot}). The favourable observation of sulfur gases in M-dwarf exoplanet atmospheres is therefore a widespread trend amongst M-dwarfs but nonetheless requires detailed photochemical modelling to be predicted due to the existence of outlier cases. The FUV\,\&\,NUV of the SEDs increases across the K-dwarfs and up to the Sun, while EUV remains low, reflecting the peak stellar emission shifting to shorter wavelengths at higher stellar effective temperature. The resulting planetary Bond albedos likewise increase with this increase in stellar UV for two reasons: the SO$_2$ photodissociation and cloud formation are greater for increasing UV flux, and the clouds that form as a result of the increased UV flux are themselves better reflectors at shorter wavelengths.

Our results for the full sample of MUSCLES SEDs help to elucidate the mapping between host star UV flux and the photochemistry of sulfur-gases in exoplanetary atmospheres (Fig.\,\ref{fig:star_plot}). Current and ongoing observations of rocky targets with JWST will soon indicate the mapping between host star EUV flux and the ability of exoplanets at different instellation fluxes to grow and maintain secondary atmospheres over geological timescales. Taken together, sulfur chemistry allows the bounds of habitability on rocky exoplanets orbiting M-dwarf host stars to be tested observationally.

A population level study over the coming years will be capable of observationally identifying surface uninhabitability and testing the inner bound of the empirical Habitability Zone, via detection of atmospheric sulfur-cycles on M-dwarf exoplanets. Performing this test will be challenging as it requires detailed photochemical modelling and appropriate stellar UV characterisation for a nuanced target selection, and must access the frontier of rocky exoplanet atmospheres with transit features on the order of 10s\,--\,100s\,ppm. It will be made easier, however, by the preponderance of potentially terrestrial Venus-Zone objects that have been detected thus far, and the general trend of observability of sulfur gases on planets hosted by M-dwarfs that we have demonstrated here. Performing this test, and constraining the inner edge of the Habitable Zone around M-dwarfs, will fundamentally shape our picture of the climate evolution of rocky worlds, and would be a triumph for the synergy between exoplanet astronomy of distant targets, and detailed planetary science at home.



\subsection*{Materials and Methods}

\subsubsection*{Photochemical-kinetics model}

We simulate the atmospheric-chemistry of the model atmospheres using the photochemical-kinetics code \textsc{Argo} and the Stand2021 chemical network \cite{Rimmer2021,Rimmer2016,Rimmer2019}. \textsc{Argo} is a 1D Lagrangian code which solves the atmospheric chemistry and photochemistry of planetary atmospheres accurately within 100\,--\,30,000\,K. The chemical network, Stand2021, is a list of reactants, products, and rate constants for every reaction considered in the model. The model treats H/C/N/O/S chemistry (plus many more elements not included in this analysis) and has been extensively validated, having previously been used to simulate the atmospheres of gas giant exoplanets \cite{Tsai2023}, sub-Neptune exoplanets \cite{Shorttle2024}, and the rocky planets Venus and Earth [e.g., \cite{Rimmer2016,Rimmer2021,Jordan2022}], agreeing closely with other commonly used photochemical-kinetics codes and chemical networks. As described in \cite{Rimmer2016} and \cite{Rimmer2021}, the chemical reactions listed in Stand2021 are solved by \textsc{Argo} at each altitude step as a set of time-dependent, coupled, non-linear differential equations:
\begin{equation}
\dfrac{dn_{\rm X}}{dt} = P_{\rm X} - L_{\rm X}n_{\rm X} - \dfrac{\partial \Phi_{\rm X}}{\partial z},
\label{eq:argo_eqn}
\end{equation}
where, at a given altitude $z\,{\rm (cm)}$ and time $t\,{\rm (s)}$, $n_{X}$ (cm$^{-3}$) is the number density of species $X$, $P_{\rm X}$ (cm$^{-3}$ s$^{-1}$) is the rate of production of species X, $L_{\rm X}$ (s$^{-1}$) is the rate constant for loss of species X, and $\partial \Phi_{\rm X}/\partial z$ (cm$^{-3}$ s$^{-1}$) describes the divergence of the vertical diffusion flux, encapsulating eddy-diffusion and molecular-diffusion.

As a Lagrangian code, \textsc{Argo} follows a parcel of gas as it rises from the surface to the top of the atmosphere and back down again. An initial condition for the chemical composition is input at the base of the atmosphere. In our analysis, the initial composition is either $\sim$\,96.5\% CO$_2$, and $\sim$\,3.5\% N$_2$, or vice versa for the N$_2$-dominated atmospheres explored in Fig.\,\ref{fig:transit_spectra}, plus trace gases. The trace gas composition is fixed at the composition observed in the atmosphere of Venus for H$_2$O (30\,ppm) and CO (20\,ppm) for simplicity. The trace gas composition of sulfur-containing gases (SO$_2$/OCS/H$_2$S) are each varied between 1\,ppm and 1\,\% as described throughout our analysis. At every step on the journey upwards \textsc{Argo} solves equation (\ref{eq:argo_eqn}) for all species present, based on the pressure, temperature, and species' abundances at that altitude. The time interval over which \textsc{Argo} solves for is prescribed by the eddy diffusion profile which parametrises vertical transport through the atmosphere. The eddy-diffusion profile for modern Venus is assumed, taken from past photochemical-kinetics models of the lower \cite{Kras2007} and middle \cite{Kras2012} atmosphere of Venus.

Volume mixing ratios of each chemical species are recorded at every grid height, constructing chemical profiles in the atmosphere. After ARGO completes one global iteration (i.e., one round trip of the parcel moving up and down again through the atmosphere) the UV radiative transfer model is run on the constructed profiles and the actinic flux throughout the atmosphere, F$_{\lambda}$(photons cm$^{-2}$ s$^{-1}$) is calculated. The actinic flux then drives photoionisation and photodissociation reactions of a given species, X, on every subsequent global iteration, via the equation:
\begin{equation}
    k_{\lambda}(X) = \frac{1}{2} \int F_{\lambda, z} \sigma_{\lambda}(X) d\lambda
\end{equation}
where $\lambda$ (\AA) is wavelength, z (km) is the altitude in the atmosphere, and $\sigma_{\lambda}$(X) (cm$^{-1}$) are the photochemical cross sections. Photoionisation and photodissociation cross sections were assembled previously from the PhIDRates \cite{HuebnerMukherjee2015} and MPI-Mainz UV/VIS \cite{Keller-Rudek2013} databases. The code and chemical network are described extensively in \cite{Rimmer2016} and \cite{Rimmer2021}.

We couple this photochemical-kinetics code with a radiative-convective model described below to compute the temperature profile of the atmospheres self-consistently. Since a pressure temperature profile is required for the atmospheric-chemistry solver, the radiative-convective model is run first, assuming constant mixing ratios of species throughout the atmosphere, fixed by the initial chemical composition. This provides the initial temperature profile for the chemistry calculation. Then, in between global runs of the photochemical-kinetics code (i.e., one full upward/downward pass through the atmosphere), the updated chemical profiles are used to inform the radiative-convective model, which recalculates the temperature structure based on the new chemical profiles. The atmospheric chemistry is therefore updated with a fixed temperature profile, and the temperature profile is updated with a fixed chemical composition, and these modules successively iterate between each other. The atmospheric chemistry code also tracks the condensation of photochemically-produced H$_2$SO$_4$, and other species not relevant to this analysis \cite{Rimmer2021}. At a given atmospheric layer, the condensation of H$_2$SO$_4$ leads to any H$_2$SO$_4$ (g) in excess of the saturation vapour pressure at that layer to be put into a reservoir of H$_2$SO$_4$ (l) which provides an effective cloud profile \cite{Rimmer2021}. While this is a simplified model of cloud formation, not accounting for the complex microphysics of aerosol nucleation, the condensed H$_2$SO$_4$ profile that results when Venus's pressure-temperature profile is used closely matches the observed altitudes of the upper and lower cloud/haze layers on Venus \cite{Jordan2021}. This effective cloud profile then provides the mass of condensed H$_2$SO$_4$ at each atmospheric layer for the climate calculation, described below. The temperature profile calculation is run to convergence in every iteration, satisfied when the system is in radiative balance with respect to both the layer-wise heating-rates within the atmosphere, and the global energy balance (i.e., absorbed stellar radiation = outgoing thermal radiation). The full coupled model iterates back and forth between the chemistry and the climate calculations until the chemical profiles converge between two successive iterations, and thus global convergence is satisfied.

\subsubsection*{Radiative-convective model}

For the climate calculation we use the line-by-line 1D radiative-convective model from \cite{Wordsworth2021}. The code calculates azimuthally averaged upwelling and downwelling spectral irradiances in a $\delta$-Eddington two-stream-type approximation. Line absorption from atmospheric gases is calculated with line coefficients from the HITRAN2020 database \cite{HITRAN2020}, accounting for CO$_2$, CO, H$_2$O, SO$_2$, OCS, and H$_2$S, and modeling the line absorption as a Voigt function. Line broadening from CO$_2$ gas is used when available, and air broadening is used if CO$_2$ broadening is not available. A line-wing cutoff of 500\,cm$^{-1}$ is used for CO$_2$ and 25\,cm$^{-1}$ for the remaining gases \cite{Wordsworth2017}. Continuum absorption due to H$_2$SO$_4$ aerosol is included, assuming a log-normal particle size distribution \cite{AckermanMarley2001} centred on a modal particle radius of 1\,$\mu$m, and a concentration of 75\,\% H$_2$SO$_4$-H$_2$O solution, using the same mass-absorption and -scattering coefficients as those used in the petitRADTRANS input opacities \cite{Molliere2019}. These values of modal particle size and acid concentration match the average of Venus's cloud deck and are assumed throughout our analysis for simplicity, and this assumption does not substantially influence any of the results that we present. Collisionally-induced absorption due to CO$_2$ and N$_2$ is included \cite{Wordsworth2017}. Convective adjustment to the dry adiabat is applied whenever the atmospheric lapse rate exceeds the dry adiabatic lapse rate, assuming Venus's surface gravity and using the heat capacity of the CO$_2$-N$_2$ background atmosphere.

The code treats longwave radiation and shortwave radiation separately. The source function in the shortwave routine is from incident radiation of the host star incoming at the top of the atmosphere and propagating through atmospheric layers. The shortwave routine accounts for absorption and scattering, detailed later on in this section. In the longwave routine, the source function is due to thermal radiation from the surface, at surface temperature T$_{s}$, and from every atmospheric layer at the given layer temperature, described by the Planck function, $B_{\nu}$:
\begin{equation}
	I_{lw, up}(\tau,\mu) = B_{\nu}(T_{s})e^{-\frac{\tau}{\mu}}+\frac{1}{\mu}\int_{0}^{\tau}B_{\nu}(\tau',\mu)e^{\frac{(\tau'-\tau)}{\mu}}d\tau'
\end{equation}
\begin{equation}
	I_{lw, dn}(\tau,\mu) = \frac{1}{\mu}\int_{\tau}^{\tau_{\infty}}B_{\nu}(\tau',\mu)e^{\frac{(\tau'-\tau)}{\mu}}d\tau'
\end{equation}
where vertical optical depth, $\tau$, at a given wavenumber, $\nu$, is defined as:
\begin{equation}
	\tau = \frac{\kappa(p_{s}-p)}{g}
\end{equation}
for pressure $p$, surface gravity $g$, and mass absorption coefficient $\kappa$. These equations are discretised and solved over evenly spaced layers in log-pressure. The mass absorption coefficients are calculated over a pressure-temperature grid, for temperature points running from 100\,K to 1000\,K at 100\,K intervals, and 30 pressure points running from 92\,$\times$\,10$^{5}$\,Pa to 1\,Pa, using log-linear interpolation to find $\kappa(p,T)$ in each atmospheric layer as the solution progresses. In the longwave routine, a layer optical depth weighting from \cite{Clough1992} is used to ensure model accuracy in the limits of both high and low absorption regions of the spectrum \cite{Wordsworth2017}.

The azimuthal averaging is discretised using 8-point Gaussian quadrature, averaged using the Gaussian weighting procedure:
\begin{equation}
	F_{up} = 2\pi \int_{0}^{1}I\mu d\mu \approx 2\pi \sum_{i} I(\tau_{i})\mu_{i} w_i
\end{equation}
for Gaussian weights $w_i$. The details of the code are described fully in \cite{Wordsworth2017} and \cite{Wordsworth2021}.

In order to account for the effects of scattering and clouds/aerosol on the planetary climate, we implement a shortwave scattering routine into the radiative-convective model described above \cite{Wordsworth2021}. The shortwave routine parametrisation describes scattering and absorption for gaseous and condensed components of the atmosphere, enabling our treatment of Venus-like atmospheres with H$_2$SO$_4$ cloud layers where aerosol scattering can have a substantial influence on spectral irradiances, planetary Bond albedo, and thus the global energy balance. The parametrisation, which effectively captures the effect of multiple scattering in cloudy atmospheres, has been applied previously in a radiative-convective model of the atmosphere of Venus \cite{Mendonca2015} and is described fully in \cite{Mendonca2015,briegleb1992}. The parametrisation first calculates the optical depth $\tau$, single scattering albedo $\omega$, asymmetry parameter $g$, and forward scattering fraction $f$, within each atmospheric layer, accounting for absorbing and scattering constituents (denoted subscript $i$), according to the equations: 
\begin{equation}
    \tau = \sum_{i}\tau_{i}
\end{equation}
\begin{equation}
    \omega = \frac{\sum_{i}\omega_{i}\tau_{i}}{\tau}
\end{equation}
\begin{equation}
    g = \frac{\sum_{i}g_{i}\omega_{i}\tau_{i}}{\omega\tau}
\end{equation}
\begin{equation}
    f = \frac{\sum_{i}f_{i}\omega_{i}\tau_{i}}{\omega\tau}
\end{equation}
These parameters are rescaled with the $\delta$ adjustment, according to the $\delta$-Eddington approximation \cite{Joseph1976}, by removing the fraction of scattered energy associated with the forward-scattered peak:
\begin{equation}
    \tau^{*} = \tau(1-\omega f)
\end{equation}
\begin{equation}
    \omega^{*} = \omega\frac{1-f}{1-\omega f}
\end{equation}
\begin{equation}
    g^{*} = \frac{g-f}{1-f}
\end{equation}
With these rescaled parameters, the fraction of spectral radiance transmitted through a layer, the \textit{transmissivity}, and the fraction reflected by a layer, the \textit{reflectivity}, can be computed for every atmospheric layer. Incoming radiation from the host star is initially direct, with angle of incidence $\mu_{o}$. Once scattered, radiation is assumed to be diffuse and isotropic, as opposed to transmitted direct radiation which continues to propagate as direct radiation with angle of incidence $\mu_{o}$. The radiation field in the atmosphere therefore has two components to it: \textit{direct} radiation, with an angular dependence, and \textit{diffuse} radiation, which is isotropic. For every layer, one thus needs to describe the transmissivity and reflectivity to \textit{direct} radiation ($T$ and $R$), and the transmissivity and reflectivity to \textit{diffuse} radiation ($\overline{T}$ and $\overline{R}$) as described in \cite{briegleb1992}:
\begin{equation}
    R(\mu_{o}) = (\alpha - \gamma)\overline{T} e^{-\frac{\tau^{*}}{\mu_{o}}} + (\alpha + \gamma)\overline{R} - (\alpha - \gamma)
\end{equation}
\begin{equation}
    T(\mu_{o}) = (\alpha - \gamma)\overline{R} e^{-\frac{\tau^{*}}{\mu_{o}}} + (\alpha + \gamma)\overline{T} - (\alpha + \gamma -1) e^{-\frac{\tau^{*}}{\mu_{o}}}
\end{equation}
\begin{equation}
    \overline{R}(\mu_{o}) = (u+1)(u-1)(e^{\lambda \tau^{*}} - e^{-\lambda\tau^{*}})N^{-1}
\end{equation}
\begin{equation}
    \overline{T}(\mu_{o}) = 4uN^{-1}
\end{equation}
for $\alpha$, $\gamma$, $\lambda$, N, and u defined as:
\begin{equation}
    \alpha = \frac{3}{4}\omega^{*}\mu_{o} \frac{1+g^{*}(1-\omega^{*})}{1-\lambda^{2}\mu_{o}^{2}}
\end{equation}
\begin{equation}
    \gamma = \frac{1}{2}\omega^{*} \frac{1+3g^{*}(1-\omega^{*})\mu_{o}^{2}}{1-\lambda^{2}\mu_{o}^{2}}
\end{equation}
\begin{equation}
    N = (u+1)^{2}e^{\lambda \tau^{*}} - (u-1)^{2}e^{-\lambda \tau^{*}}
\end{equation}
\begin{equation}
    u = \frac{3(1-\omega^{*}g^{*})}{2\lambda}
\end{equation}
\begin{equation}
    \lambda = \sqrt{3(1-\omega^{*})(1-\omega^{*}g^{*})}
\end{equation}
The reflectivity and transmissivity to direct and diffuse radiation in a given layer can then be used to calculate the reflectivity and transmissivity to direct and diffuse radiation at every layer interface using the `adding-layer' method \cite{briegleb1992}. The adding-layer method makes two passes through the atmosphere on every iteration: on the downwards pass, moving from the top of the atmosphere downwards, reflectivity and transmissivity of layers are subsequently combined with the reflectivity and transmissivity of the entire column above until the surface is reached; on the upwards pass, layers are combined subsequently with the column below until the top of the atmosphere is reached. The combination of two overlying layers (or a layer combined to the column below or above), with layer 1 overlying layer 2, is calculated as:
\begin{equation}
    R_{12}(\mu_{o}) = R_{1}(\mu_{o})+\frac{\overline{T}_{1}((T_{1}(\mu_{o})-e^{-\frac{\tau_{1}^{*}}{\mu_{o}}})\overline{R}_{2} + e^{-\frac{\tau_{1}^{*}}{\mu_{o}}}R_{2}(\mu_{o}))}{1-\overline{R}_{1}\overline{R}_{2}}
\end{equation}
\begin{equation}
	T_{12}(\mu_{o}) = e^{-\frac{\tau_{1}^{*}}{\mu_{o}}}T_{2}(\mu_{o}) + \frac{\overline{T}_{2}((T_{1}(\mu_{o})-e^{-\frac{\tau_{1}^{*}}{\mu_{o}}}) + e^{-\frac{\tau_{1}^{*}}{\mu_{o}}}R_{2}(\mu_{o})\overline{R}_{1})}{1-\overline{R}_{1}\overline{R}_{2}}
\end{equation}
\begin{equation}
	\overline{R}_{12} = \overline{R}_{1} + \frac{\overline{T}_{1}\overline{R}_{2}\overline{T}_{1}}{1-\overline{R}_{1}\overline{R}_{2}}
\end{equation}
\begin{equation}
	\overline{T}_{12} = \frac{\overline{T}_{1}\overline{T}_{2}}{1-\overline{R}_{1}\overline{R}_{2}}
\end{equation}
The formulas above provide the combined reflectivities and transmissivities for upwelling and downwelling radiation at every layer interface. At a layer interface, with rescaled optical depth $\tau^{*}$ from the top of the atmosphere to the interface: $e^{-\frac{\tau^{*}}{\mu_{o}}}$ describes the direct beam transmission from the top of the atmosphere the the interface; $R_{up}(\mu_{o})$ describes the reflectivity of the entire column below the interface to direct radiation incident from above; $T_{dn}(\mu_{o})$ describes the total transmission of the entire column above the interface to radiation incident from above; $\overline{R}_{up}$ describes the reflectivity of the entire column below the interface to diffuse radiation from above; $\overline{R}_{dn}$ describes the reflectivity of the entire column above the interface to diffuse radiation from below. The upwelling and downwelling spectral fluxes can then be evaluated at every layer interface according to:
\begin{equation}
	F_{up} = \frac{e^{-\frac{\tau^{*}}{\mu_{o}}}R_{up}(\mu_{o}) + (T_{dn}(\mu_{o})-e^{-\frac{\tau^{*}}{\mu_{o}}})\overline{R}_{up}}{1-\overline{R}_{dn}\overline{R}_{up}}
\end{equation}
\begin{equation}
	F_{dn} = e^{-\frac{\tau^{*}}{\mu_{o}}} + \frac{(T_{dn}(\mu_{o})-e^{-\frac{\tau^{*}}{\mu_{o}}}) + e^{-\frac{\tau^{*}}{\mu_{o}}}R_{up}(\mu_{o})\overline{R}_{dn}}{1-\overline{R}_{dn}\overline{R}_{up}}
\end{equation}
The resulting upward and downward spectral fluxes are summed to obtain spectrally integrated fluxes, and then differenced to obtained net layer-wise heating rates.



\clearpage 

%

%
%
%
%
%
%


\section*{Acknowledgments}
The authors thank David Wilson and Greg Cooke for helpful conversations about the Trappist-1 SED. 
\paragraph*{Funding:}
S.J. thanks the Science and Technology Facilities Council (STFC) for the PhD studentship (grant reference ST/V50659X/1).
\paragraph*{Author contributions:}
S.J. conceived of the study and implemented the simulations. S.J. and O.S. wrote and edited the manuscript. P.B.R. created the photochemistry code and contributed to editing of the manuscript.
\paragraph*{Competing interests:} There are no competing interests to declare.
\paragraph*{Data and materials availability:}
All data needed to evaluate the conclusions in the paper are present in the paper and/or the Supplementary Materials. This study made use of the software petitRADTRANS for simulating transit spectra of exoplanets \cite{Molliere2019,Molliere2020,Alei2022}, PCM\_LBL for simulating planetary pressure-temperature profiles \cite{Wordsworth2021}, and the photochemical-kinetics code ARGO and chemical network STAND for simulating atmospheric chemistry \cite{Rimmer2016,Rimmer2019,Rimmer2021}. petitRADTRANS is available online at (\url{https://petitradtrans.readthedocs.io/en/latest/}), PCM\_LBL is available on github (\url{https://github.com/wordsworthgroup/mars_redox_2021/tree/main/PCM_LBL}), and ARGO and STAND are described in the publicly available set of papers \cite{Rimmer2016,Rimmer2019,Rimmer2021}. We also make use of data from the HITRAN2020 database available online at (\url{https://hitran.org/data-index/}) and the MUSCLES \cite{France2016, Youngblood2016, Loyd2016} and Mega-MUSCLES \cite{Wilson2021} databases available online at (\url{https://archive.stsci.edu/prepds/muscles/}).


\subsection*{Supplementary materials}
Supplementary Discussion\\


\newpage


\renewcommand{\thefigure}{S\arabic{figure}}
\renewcommand{\thetable}{S\arabic{table}}
\renewcommand{\theequation}{S\arabic{equation}}
\renewcommand{\thepage}{S\arabic{page}}
\setcounter{figure}{0}
\setcounter{table}{0}
\setcounter{equation}{0}
\setcounter{page}{1} 


\begin{center}
\section*{Supplementary Materials for\\ \scititle}

Sean~Jordan$^{\ast}$,
Oliver~Shorttle,
Paul~B.~Rimmer\\ 
\small$^\ast$Corresponding author. Email: jordans@ethz.ch\\
\end{center}

\subsubsection*{This PDF file includes:}
Supplementary Discussion

\newpage


\subsection*{Supplementary Discussion}

\subsubsection*{Implications of observing OCS or H$_2$S}

On Venus, SO$_2$ is observed in the deep atmosphere with a mixing ratio around $\sim$\,150\,ppm. The other sulfur gases, OCS and H$_2$S, have been observed \textit{in situ}, alongside SO$_2$, at abundances of order $\sim$\,1\,--\,10\,ppm in the deep atmosphere \cite{Rimmer2021}. These gases are efficiently photochemically destroyed on Venus before they can reach the upper atmosphere where remote sensing could detect their presence. Our results demonstrate that OCS and H$_2$S are also sensitive to the widely ranging UV fluxes from different M-dwarf host stars. For the low-UV case, OCS and H$_2$S remain observable in the upper atmosphere, and have spectroscopic features that can be accessed by JWST.

Linking the presence of atmospheric OCS or H$_2$S to a diagnosis of surface conditions on an exoplanet has not been as well investigated as the presence of atmospheric SO$_2$ and H$_2$SO$_4$ \cite{Loftus2019}. SO$_2$ and H$_2$SO$_4$ participate in the oxidised sulfur cycle where sulfur has oxidation state +4 in SO$_2$ and +6 in H$_2$SO$_4$. The sulfur atom in OCS on the other hand has oxidation state 0 and in H$_2$S has oxidation state -2. The direct connection between the presence of surface water and the reduced sulfur cycle has not yet been well studied and our results suggest that this deserves future investigation. Linking the presence of atmospheric OCS or H$_2$S to the hydrological cycle of a planet would require investigating the solubility equilibria of OCS and H$_2$S, and their aqueous chemistry. If the aqueous products of dissolved OCS or H$_2$S can precipitate out of solution into mineral phases then these sulfur gases may also serve as useful uninhabitability indicators via wet deposition \cite{Loftus2019}.

A separate observational implication of the observability of SO$_2$, OCS and H$_2$S in the atmospheres of M-dwarf exoplanets is that the relative abundances of each gas could be constrained simultaneously. For low-UV M-dwarf targets, constraining the relative abundance of oxidised and reduced sulfur gases will open a window into a planet's interior oxygen fugacity [e.g., \cite{liggins2022}], provided that there is sufficient sulfur-outgassing at the surface. If sulfur gases are supplied to the atmosphere via volcanic degassing, then the relative speciation of atmospheric sulfur will depend on the oxygen fugacity of the mantle source from which the melt was derived \cite{liggins2022, liggins2023, Guimond2023}. For low-UV M-dwarf exoplanets, the relative proportions of SO$_2$:OCS:H$_2$S are effectively undepleted by photochemistry, and so the observed ratio will equal their real ratio in the deep atmosphere. Our result therefore demonstrate that sulfur chemistry can provide a probe of the oxygen fugacity of a planet's mantle around a low-UV M-dwarf host star, which will be accessible with transmission spectroscopy.

\subsubsection*{Alternative sources and sinks of sulfur}

The prospect of observing spectroscopic features of SO$_2$, OCS, or H$_2$S in an exoplanet's atmosphere each have potential caveats, inspired by unknown processes in Venus's sulfur cycle. First, the efficiency of cloud formation in photochemical models of Venus is not sufficient to reproduce the observations of SO$_2$, H$_2$O, and H$_2$SO$_4$ in the clouds and upper atmosphere [e.g., \cite{BiersonZhang2020,Rimmer2021}]. In order to reproduce observations, an additional source of hydrogen atoms is required in combination with chemistry beyond only gas phase interactions so that SO$_2$ can be more efficiently sequestered in the clouds. Previous work has demonstrated how this could be achieved with aqueous chemistry inside the cloud droplets and a source of mineral dust delivered to the clouds \cite{Rimmer2021}. Alternatively, the observed profile of SO$_2$ on Venus can be reproduced with the speculative biochemistry of cloud-based microorganisms in the cloud droplets if there is an additional supply of reducing gases in the deep atmosphere \cite{Jordan2022, Bainsammonia}. Upcoming missions to Venus will be able to confirm or deny these hypotheses, however currently these processes cannot be generalised to exoplanets. Due to this limitation, we do not prescribe any additional chemistry beyond the standard gas-phase chemical network. The enhanced depletion of SO$_2$ that is observed in the clouds of Venus therefore implies that the survival of self-shielding SO$_2$ above the cloud-layers of exoplanets may be different to the results that we have found. This has the potential to influence the abundances of deep atmosphere SO$_2$ that are required before above-cloud SO$_2$ signatures could prevail on targets irradiated by a high-UV M-dwarf that may be accessible with next-generation observatories.

Second is the unknown thermochemical pathway converting some OCS to CO below the cloud base of Venus. Between $\sim$\,30\,--\,40\,km altitude, the OCS mixing ratio is observed to decrease from $\sim$\,40\,ppm to $\sim$\,1\,ppm. Within this altitude range, the CO mixing ratio is observed to increase proportionally, leading to the conclusion that there is unknown OCS chemistry responsible for the conversion to CO and a sulfur-containing product such as a sulfur allotrope. Since the chemical reactions involved remain unknown this cannot be predicted for the case of exoplanets and therefore the observability of OCS in the upper atmosphere is subject to uncertainties on its behaviour in the deep atmosphere. Future work constraining the pressure and temperature dependence of possible destruction pathways for OCS in the deep atmosphere of Venus will provide valuable insight into the observability of OCS on low-UV M-dwarf exoplanets. The possible destruction pathways of OCS may be possible to verify \textit{in situ} with the upcoming missions exploring Venus's deep atmosphere chemistry \cite{Garvin2022}.

Third, there remains the possibility that sulfur-metabolising life could exist in the cloud layer of Venus. Sulfur-metabolising life could catalyse conversions between sulfur-gases based on known sulfur-based energy metabolisms of terrestrial microorganisms \cite{SchulzeMakuch2004,SchulzeMakuch2006}. These metabolic pathways have been thoroughly investigated in the context of Venus's atmosphere previously \cite{Jordan2022} which revealed that SO$_2$ could be converted to OCS, H$_2$S, or sulfur allotropes via the biochemical action of life in a H$_2$SO$_4$ cloud layer, if such extreme life is possible. On Venus, there has not yet been enough reducing power observed in the deep atmosphere for this conversion to be taking place and thus an abundant biosphere, shaping the atmospheric sulfur chemistry, was ruled out \cite{Jordan2022}. However, on sulfur-rich exoplanets this may not necessarily be the case, which therefore poses an alternative mechanism for the scrubbing of SO$_2$ out of the upper atmosphere and mimicking the presence of surface water oceans. Conversely, if other biosignatures could be detected in combination with this inference then sulfur-metabolising life could be discovered in aerial biospheres in the cloud layers of canonically-uninhabitable exoplanets. This will be a profound avenue of future research if the upcoming Morning Star missions to Venus discover evidence of stable biochemistry in the sulfuric acid cloud droplets \cite{Seager2022}.

\subsubsection*{Stellar UV variability.}

One additional caveat to our results is that M-dwarfs can often exhibit energetic flaring and coronal mass ejection (CME) \cite{Loyd2018}. Flares and CMEs are transient events and therefore will influence the steady state background atmospheres which we have restricted ourselves to in this work. If, however, flaring events are sufficiently frequent and energetic then this has the potential to drive different sulfur photochemistry in an exoplanet's atmosphere. Constraining the regions of parameter space where flaring frequency and flaring energy could preclude observations of sulfur chemistry in an exoplanet's atmosphere would be a valuable avenue of future research. We note also that future observations, which could diagnose the photochemical effect of stellar flaring on exoplanet atmospheres, will nonetheless require knowing the chemistry of the background atmosphere that results under quiescence. Our investigation into the sensitivity of sulfur photochemistry may therefore also be applicable for measuring flare-driven photochemistry around transiently active host stars.



\end{document}